\documentstyle[11pt,aaspp4]{article}
\begin{document}

\title{DIRECT Distances to Nearby Galaxies Using Detached Eclipsing
Binaries and Cepheids. V. Variables in the Field M31F\footnote{Based
on the observations collected at the F.~L.~Whipple Observatory (FLWO)
1.2~m telescope and at the Michigan-Dartmouth-MIT (MDM)
1.3~m telescope}}

\author{B. J. Mochejska and J. Kaluzny}
\affil{Warsaw University Observatory, Al. Ujazdowskie 4,
PL-00-478 Warszawa, Poland} 
\affil{\tt e-mail: mochejsk@sirius.astrouw.edu.pl, jka@sirius.astrouw.edu.pl} 
\author{K. Z. Stanek\altaffilmark{2}, M. Krockenberger and D. D. Sasselov } 
\affil{Harvard-Smithsonian Center for Astrophysics, 60 Garden St.,
Cambridge, MA~02138} 
\affil{\tt e-mail: kstanek@cfa.harvard.edu, krocken@cfa.harvard.edu, 
sasselov@cfa.harvard.edu} 
\altaffiltext{2}{On leave from N.~Copernicus Astronomical Center, 
Bartycka 18, Warszawa PL-00-716, Poland} 

\begin{abstract}

We undertook a long term project, DIRECT, to obtain the direct
distances to two important galaxies in the cosmological distance
ladder -- M31 and M33 -- using detached eclipsing binaries (DEBs) and
Cepheids. While rare and difficult to detect, DEBs provide us with the
potential to determine these distances with an accuracy better than
5\%. The extensive photometry obtained in order to detect DEBs
provides us with good light curves for the Cepheid variables. These
are essential to the parallel project to derive direct Baade-Wesselink
distances to Cepheids in M31 and M33. For both Cepheids and eclipsing
binaries, the distance estimates will be free of any intermediate
steps.
 
As a first step in the DIRECT project, between September 1996 and
October 1997 we obtained 95 full/partial nights on the F. L. Whipple
Observatory 1.2~m telescope and 36 full nights on the
Michigan-Dartmouth-MIT 1.3~m telescope to search for DEBs and new
Cepheids in the M31 and M33 galaxies.  In this paper, fifth in the
series, we present the catalog of variable stars found in the field M31F
$[(\alpha,\delta)= (10.\!\!\arcdeg10, 40.\!\!\arcdeg72), {\rm J2000.0}]$.
We have found 64 variable stars: 4 eclipsing binaries, 52 Cepheids and 8
other periodic, possible long period or non-periodic variables. The catalog
of variables, as well as their photometry and finding charts, is
available via {\tt anonymous ftp} and the {\tt World Wide Web}.  The
complete set of the CCD frames is available upon request.

\end{abstract}


\section{Introduction}

Starting in 1996 we undertook a long term project, DIRECT (as in
``direct distances''), to obtain the distances to two important
galaxies in the cosmological distance ladder -- M31 and M33 -- using
detached eclipsing binaries (DEBs) and Cepheids.  These two nearby
galaxies are stepping stones to most of our current effort to
understand the evolving universe at large scales.  First, they are
essential to the calibration of the extragalactic distance scale
(Jacoby et al.~1992; Tonry et al.~1997). Second, they constrain
population synthesis models for early galaxy formation and evolution
and provide the stellar luminosity calibration. There is one simple
requirement for all this---accurate distances.

DEBs have the potential to establish distances to M31 and M33 with an
unprecedented accuracy of better than 5\% and possibly to better than
1\%. These distances are now known to no better than 10-15\%, as there
are discrepancies of $0.2-0.3\;{\rm mag}$ between various distance
indicators (e.g.~Huterer, Sasselov \& Schechter 1995; Holland 1998;
Stanek \& Garnavich 1998).  Detached eclipsing binaries (for reviews
see Andersen 1991; Paczy\'nski 1997) offer a single step distance
determination to nearby galaxies and may therefore provide an accurate
zero point calibration---a major step towards very accurate
determination of the Hubble constant, presently an important but
daunting problem for astrophysicists. A DEB system was recently used
by Guinan et al.~(1998) and Udalski et al.~(1998) to obtain an
accurate distance estimate to the Large Magellanic Cloud.

The detached eclipsing binaries have yet to be used (Huterer et
al.~1995; Hilditch 1996) as distance indicators to M31 and M33.
According to Hilditch (1996), there were about 60 eclipsing binaries
of all kinds known in M31 (Gaposchkin 1962; Baade \& Swope 1963, 1965)
and only {\em one} in M33 (Hubble 1929), none of them observed with
CCDs.  Only now does the availability of large-format CCD detectors
and inexpensive CPUs make it possible to organize a massive search for
periodic variables, which will produce a handful of good DEB
candidates. These can then be spectroscopically followed-up with the
powerful new 6.5-10 meter telescopes.

The study of Cepheids in M31 and M33 has a venerable history (Hubble
1926, 1929; Gaposchkin 1962; Baade \& Swope 1963, 1965).  Freedman \&
Madore (1990) and Freedman, Wilson \& Madore (1991) obtained
multi-band CCD photometry of some of the already known Cepheids, to
build period-luminosity relations in M31 and M33, respectively.
However, both the sparse photometry and the small samples (11 Cepheids
in M33 and 38 Cepheids in M31) do not provide a good basis for
obtaining direct Baade-Wesselink distances (see, e.g., Krockenberger,
Sasselov \& Noyes 1997) to Cepheids---the need for new digital
photometry has been long overdue. Recently, Magnier et al.~(1997)
surveyed large portions of M31, which have previously been ignored,
and found some 130 new Cepheid variable candidates.  Their light
curves are, however, rather sparsely sampled and in the $V$-band only.

In Kaluzny et al.~(1998, 1999, hereafter: Papers I and IV) 
and Stanek et al.~(1998, 1999, hereafter: Papers II and III), 
the first four papers of the series, we presented the catalogs 
of variable stars found in four fields in M31, called M31B, M31A, 
M31C and M31D. Here we present the catalog
of variables from the field M31F. In Sec.2 we discuss the
selection of the fields in M31 and the observations. In Sec.3 we
describe the data reduction and calibration. In Sec.4 we discuss
briefly the automatic selection we used for finding the variable
stars. In Sec.5 we discuss the classification of the variables.  In
Sec.6 we present the catalog of variable stars, followed by a brief
discussion of the results in Sec.7.

\section{Fields selection and observations}

M31 was primarily observed in 1996 with the 1.3~m McGraw-Hill
Telescope at the Michigan-Dartmouth-MIT (MDM) Observatory. We used the
front-illuminated, Loral $2048^2$ CCD ``Wilbur'' (Metzger, Tonry \&
Luppino 1993), which at the $f/7.5$ station of the 1.3~m telescope has
a pixel scale of $0.32\; arcsec\; pixel^{-1}$ and field of view of
roughly $11\;arcmin$. We used Kitt Peak Johnson-Cousins $BVI$ filters.
Data for M31 were also obtained, mostly in 1997, with the 1.2~m
telescope at the F. L. Whipple Observatory (FLWO), where we used
``AndyCam'' (Szentgyorgyi et al.~1999), with a thinned,
back-illuminated, AR coated Loral $2048^2$ pixel CCD.  The pixel scale
happens to be essentially the same as at the MDM 1.3~m telescope. We
used standard Johnson-Cousins $BVI$ filters.

Fields in M31 were selected using the MIT photometric survey of M31 by
Magnier et al.~(1992) and Haiman et al.~(1994) (see Paper I, Fig.1).
We selected six $11'\times11'$ fields, M31A--F, four of them (A--D)
concentrated on the rich spiral arm in the northeast part of M31, one
(E) coinciding with the region of M31 searched for microlensing by
Crotts \& Tomaney (1996), and one (F) containing the giant star
formation region known as NGC206 (observed by Baade \& Swope
1965). Fields A--C were observed during September and October 1996
five to eight times per night in the $V$ band, resulting in total of
110--160 $V$ exposures per field. Fields D--F were observed once a
night in the $V$-band. Some exposures in $B$ and $I$ were also
taken. M31 was also observed, in 1996 and 1997, at the FLWO 1.2~m
telescope, whose main target was M33.

In this paper we present the results for the M31F field.  We obtained
for this field useful data during 29 nights at the MDM, collecting a
total of $28\times 900\;sec$ exposures in $V$ and $2\times 600\;sec$
exposures in $I$. We also obtained for this field useful data during
22 nights at the FLWO, in 1996 and 1997, collecting a total of
$80\times 900\;sec$ exposures in $V$, $67\times 600\;sec$ exposures in
$I$ and $7\times 1200\;sec$ exposures of $B$.\footnote{The complete
list of exposures for this field and related data files are available
through {\tt anonymous ftp} on {\tt cfa-ftp.harvard.edu}, in {\tt
pub/kstanek/DIRECT} directory. Please retrieve the {\tt README} file
for instructions.  Additional information on the DIRECT project is
available through the {\tt WWW} at {\tt
http://cfa-www.harvard.edu/\~\/kstanek/DIRECT/}.}

\section{Data reduction, calibration and astrometry}

The details of the reduction procedure were given in Paper I.
Preliminary processing of the CCD frames was done with the standard
routines in the IRAF-CCDPROC package.\footnote{IRAF is distributed by
the National Optical Astronomy Observatories, which are operated by
the Associations of Universities for Research in Astronomy, Inc.,
under cooperative agreement with the NSF} Stellar profile photometry
was extracted using the {\it Daophot/Allstar} package (Stetson 1987,
1992).  We selected a ``template'' frame for each filter using a
single frame of particularly good quality.  These template images were
reduced in a standard way (Paper I).  Other images were reduced using
{\it Allstar} in the fixed-position-mode using as an input the
transformed object list from the template frames.  For each frame the
list of instrumental photometry derived for a given frame was
transformed to the common instrumental system of the appropriate
``template'' image.  Photometry obtained for the $B,V$ and $I$ filters
was combined into separate data bases. M31F $I$-band images obtained at the
MDM were reduced using FLWO ``templates''. Two templates were used in the
case of $V$-band images. An MDM template was used to fix the positions of
the stars. The $V$ photometry was transformed to the instrumental
system of an FLWO template. 

The photometric $VI$ calibration of the MDM data was discussed in
Paper I. In addition, for the field M31F on the night of 1997 October
9/10 we have obtained independent $BVI$ calibration with the FLWO
1.2~m telescope. There was an offset of $-0.020\;{\rm mag}$ in $V$ and
$0.057\;{\rm mag}$ in $V-I$ between the FLWO and the MDM calibration.
The $V$ offset is well within our estimate of the total $0.05\;mag$
systematic error discussed in Paper I, and the $V-I$ offset falls slightly
above.

We also derived equatorial coordinates for all objects included in the
data bases for the $V$ filter. The transformation from rectangular
coordinates to equatorial coordinates was derived using 79 stars
identified in the USNO-A2.0 catalog. We have also compared these
coordinates with those given by Magnier et al.~(1992) and find a
good agreement, with an average difference of $0.5\; arcsec$ 
for 66 of the transformation stars found in both catalogs. 

\section{Selection of variables}

\begin{figure}[t]
\plotfiddle{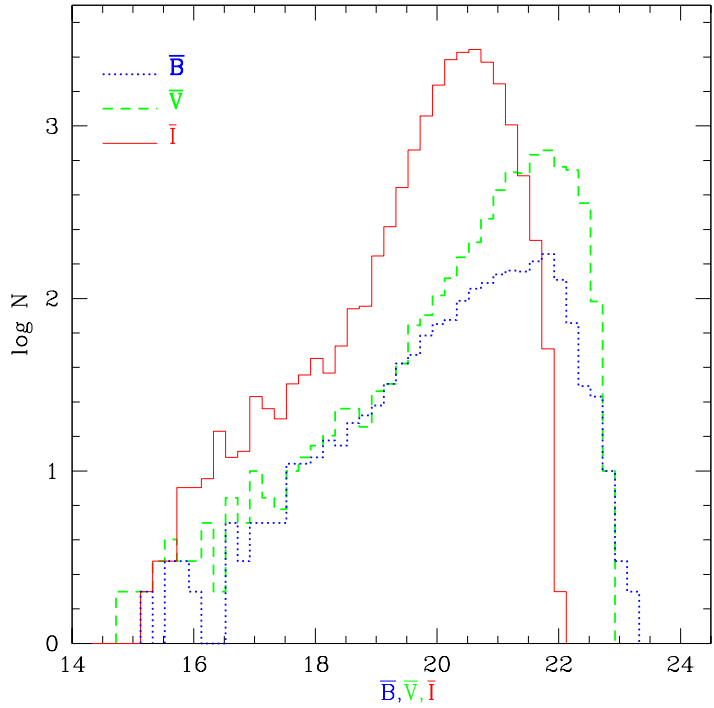}{8cm}{0}{50}{50}{-160}{-85}
\caption{Distributions in $B$ (dotted line), $V$ (dashed line) and $I$
(continuous line) of stars in the field M31F.}
\label{fig:dist}
\end{figure}

The procedure for selecting the variables was described in detail in
Paper I, so here we only give a short description, noting changes when
necessary.  The reduction procedure described in previous section
produces databases of calibrated $BVI$ magnitudes and their standard
errors. The $V$ database for M31F field contains 7997 stars, with up
to 108 measurements, the $I$ database contains 26540 stars with up to
69 measurements and the $B$ database contains 3382 stars with up to 7
measurements.  Figure~\ref{fig:dist} shows the distributions of stars
as a function of mean $\bar{B}$, $\bar{V}$ or $\bar{I}$ magnitude.  As
can be seen from the shape of the histograms, our completeness starts
to drop rapidly at about $\bar{B}\sim22$, $\bar{V}\sim22$ and
$\bar{I}\sim20.5$. The primary reason for this difference in the depth
of the photometry between $BV$ and $I$ is the level of the combined
sky and background light, which is about three times higher in the $I$
filter than in the $BV$ filters.

The measurements flagged as ``bad'' (with unusually large {\em
Daophot} errors, compared to other stars) and measurements with errors
exceeding the average error, for a given star, by more than $4\sigma$
are removed.  Usually zero to 10 points are removed, leaving the
majority of stars with roughly $N_{good}\sim95-105$ $V$\/
measurements.  For further analysis we use only those stars that have
at least $N_{good}>N_{max}/2\;(=54)$ measurements. There are 5838
such stars in the $V$ database of the M31F field.

\begin{figure}[t]
\plotfiddle{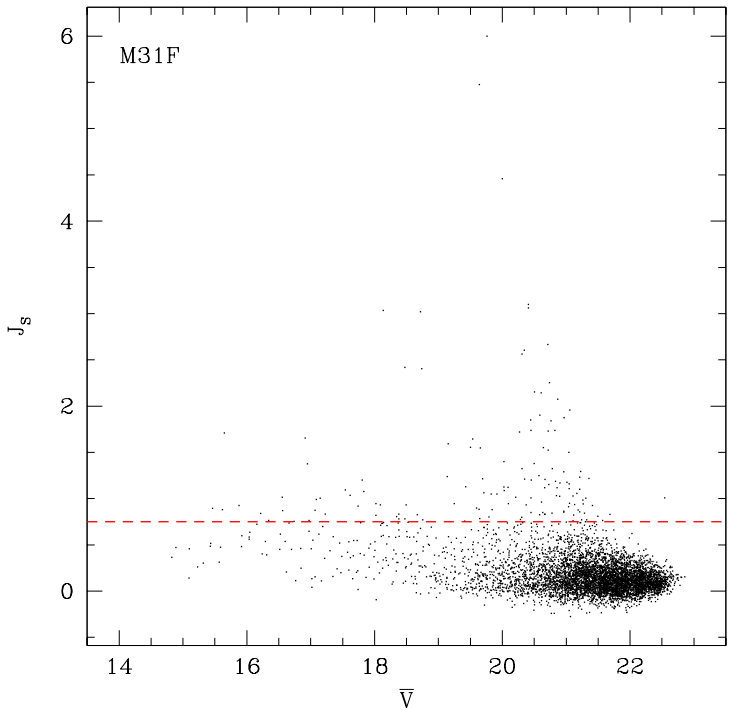}{8cm}{0}{50}{50}{-160}{-85}
\caption{Variability index $J_S$ vs. mean $\bar{V}$ magnitude for 5838
stars in the field M31F with $N_{good}>54$.  Dashed line at $J_S=0.75$
defines the cutoff applied for variability.}
\label{fig:stetj}
\end{figure}

Our next goal is to select a sample of variable stars from the total
sample defined above. There are many ways to proceed, and we largely
follow the approach of Stetson (1996), also described in Paper~I.  In
short, for each star we compute the Stetson's variability index $J_S$
(Paper I, Eq.7), and stars with values exceeding some minimum value
$J_{S,min}$ are considered candidate variables.  The definition of
$J_S$ is rooted in the assumption that on each visit to the program
field at least one pair of observations is obtained, and only when
both observations have the residual from the mean of the same sign
does the pair contribute positively to the variability index.  The
definition of Stetson's variability index includes the standard errors
of individual observations.  If, for some reason, these errors were
over- or underestimated, we would either miss real variables, or
select spurious variables as real ones. Using the procedure described
in Paper I, we scale the {\em Daophot} errors to better represent the
``true'' photometric errors.  We then select the candidate variable
stars by computing the value of $J_S$ for the stars in our $V$
database.  We used a cutoff of $J_{S,min}=0.75$ and additional cuts
described in Paper I to select 122 candidate variable stars (about 2\% of
the total number of 5838).  In Figure~\ref{fig:stetj} we plot the
variability index $J_S$ vs. apparent visual magnitude $\bar{V}$ for 5838
stars with $N_{good}>54$. 

\section{Period determination, classification of variables}

We based our candidate variables selection on the $V$ band data collected
at the MDM and the FLWO telescopes. We also have the $BI$-bands data for
the field, up to 69 $I$-band epochs and up to 7 $B$-band epochs, although
for a variety of reasons some of the candidate variable stars do not have a
$B$ or $I$-band counterpart. We will therefore not use the $BI$ data for
the period determination and broad classification of the variables. We will 
however use the $BI$ data for the ``final'' classification of some
variables.

Next we searched for the periodicities for all 122 candidate
variables, using a variant of the Lafler-Kinman (1965) string-length
technique proposed by Stetson (1996). Starting with the minimum period
of $0.25\;days$, successive trial periods are chosen so
\begin{equation}
P_{j+1}^{-1}=P_{j}^{-1}-\frac{0.02}{\Delta t},
\end{equation}
where $\Delta t=t_{N}-t_{1}=398\;days$ is the time span of the series.
The maximum period considered is $150\;days$.  For each candidate
variable 10 best trial periods are selected (Paper I) and then used in
our classification scheme.

The variables we are most interested in are Cepheids and eclipsing
binaries (EBs). We therefore searched our sample of variable stars for
these two classes of variables. As mentioned before, for the broad
classification of variables we restricted ourselves to the $V$ band
data.  We will, however, present and use the $BI$-bands data, when
available, when discussing some of the individual variable stars.

For EBs, we used the search strategy described in Paper II.  Within our
assumption the light curve of an EB is determined by nine parameters:
the period, the zero point of the phase, the eccentricity, the
longitude of periastron, the radii of the two stars relative to the
binary separation, the inclination angle, the fraction of light coming
from the bigger star and the uneclipsed magnitude. A total of six
variables passed all of the criteria. We then went back to the CCD
frames and tried to see by eye if the inferred variability is indeed
there, especially in cases when the light curve is very noisy/chaotic.
We decided to remove two dubious eclipsing binaries, classifying one as a
periodic variable with half the determined period. Its light curve is
presented in Section~6.3. The remaining four EBs with their parameters and
light curves are presented in the Section~6.1.

In the search for Cepheids we followed the approach by Stetson (1996)
of fitting template light curves to the data. We used the
parameterization of Cepheid light curves in the $V$-band as given by
Stetson (1996). There was a total of 64 variables passing all of the
criteria (Paper I and II), but after investigating the CCD frames we
removed 12 dubious ``Cepheids'', which leaves us with 52 probable
Cepheids. Their parameters and light curves are presented in 
Section~6.2.

After the selection of four eclipsing binaries, 52 Cepheids and one
periodic variable, we were left with 65 ``other'' variable stars.
After raising the threshold of the variability index to
$J_{S,min}=1.2$ (Paper I) we are left with 17 variables. After
investigating the CCD frames we removed 10 dubious variables from the
sample, which leaves seven variables which we classify as
miscellaneous. Their parameters and light curves are presented in the
Section~6.4.

\section{Catalog of variables}

In this section we present light curves and some discussion of the 64
variable stars discovered by our survey in the field M31F.
\footnote{Complete $V$ and (when available) $BI$ photometry and
$128\times128\;pixel$ ($\sim 40''\times40''$) $V$ finding charts for
all variables are available from the authors via the {\tt anonymous
ftp} from the Harvard-Smithsonian Center for Astrophysics and can be
also accessed through the {\tt World Wide Web}.}  The variable stars
are named according to the following convention: letter V for
``variable'', the number of the star in the $V$ database, then the
letter ``D'' for our project, DIRECT, followed by the name of the
field, in this case (M)31F, e.g. V244 D31F.  Tables~\ref{table:ecl},
\ref{table:ceph}, \ref{table:per} and \ref{table:misc} list the
variable stars sorted broadly by four categories: eclipsing binaries,
Cepheids, other periodic variables and ``miscellaneous'' variables, in
our case meaning ``variables with no clear periodicity''. 

\subsection{Eclipsing binaries}

\begin{figure}[p]
\plotfiddle{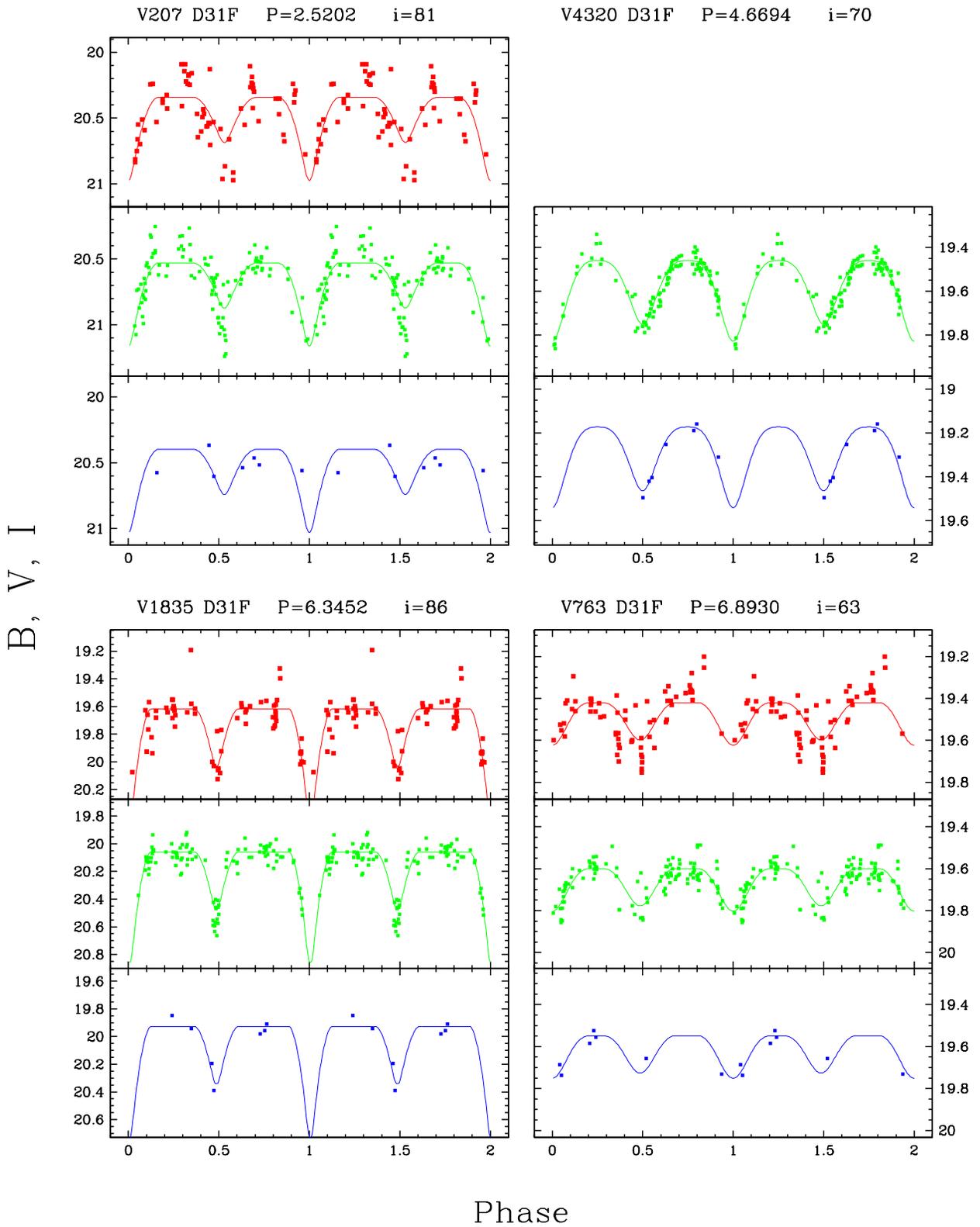}{19.5cm}{0}{83}{83}{-260}{-40}
\caption{$BVI$ lightcurves of eclipsing binaries found in the field
M31F. The thin continuous line represents the best fit model for each
star and photometric band. $B$-band lightcurve is shown in the bottom
panel and $I$-band lightcurve (when present) is shown in the top
panel.}
\label{fig:ecl}
\end{figure}

In Table~\ref{table:ecl} we present the parameters of the four eclipsing
binaries in the M31F field.  The light curves of these variables are
shown in Figure~\ref{fig:ecl}, along with the simple eclipsing binary
models discussed in the Papers I and II.  The variables are sorted in the 
Table~\ref{table:ecl} by the increasing value of the period $P$. For
each eclipsing binary we present its name, J2000.0 coordinates (in
degrees), period $P$, magnitudes $V_{max}, I_{max}$ and $B_{max}$ of
the system outside of the eclipse, and the radii of the binary
components $R_1,\;R_2$ in the units of the orbital separation.  We
also give the inclination angle of the binary orbit to the line of
sight $i$ and the eccentricity of the orbit $e$. The reader should
bear in mind that the values of $V_{max},\;I_{max},\;B_{max},\;
R_1,\;R_2,\;i$ and $e$ are derived with a straightforward model of the
eclipsing system, so they should be treated only as reasonable
estimates of the ``true'' value.

One of the eclipsing binaries found, V1835 D31F, is a good DEB candidate.
However, a much better light curve is necessary to accurately establish the
properties of the system.

\begin{small}
\tablenum{1}
\begin{planotable}{lrrrrrrrrcrl}
\tablewidth{40pc}
\tablecaption{\sc DIRECT Eclipsing Binaries in M31F}
\tablehead{ \colhead{Name} & \colhead{$\alpha_{J2000.0}$} &
\colhead{$\delta_{J2000.0}$} & \colhead{$P$} & \colhead{} & \colhead{} &
\colhead{} & \colhead{} & \colhead{} & \colhead{$i$} & \colhead{} &
\colhead{} \\  
\colhead{(D31F)} & \colhead{$(\deg)$} & \colhead{$(\deg)$}
& \colhead{$(days)$} & \colhead{$V_{max}$}  & \colhead{$I_{max}$}
& \colhead{$B_{max}$} & \colhead{$R_1$} & \colhead{$R_2$} &
\colhead{(deg)} & \colhead{$e$} & \colhead{Comments}}
\startdata  
 V207\ldots   & 10.2157 & 40.6399 &  2.5202 & 20.53 & 20.34 & 20.40 & 0.48 & 0.39 & 81 & 0.05 & \nl 
V4320\dotfill & 10.1076 & 40.7456 &  4.6694 & 19.46 & \nodata & 19.17 & 0.58 & 0.42 & 70 & 0.00 & \nl 
V1835\dotfill & 10.1433 & 40.7184 &  6.3452 & 20.06 & 19.62 & 19.93 & 0.37 & 0.30 & 86 & 0.03 & DEB\nl 
 V763\dotfill & 10.1807 & 40.7263 &  6.8930 & 19.60 & 19.42 & 19.55 & 0.64 & 0.31 & 63 & 0.02 & W UMa\nl 
\enddata
\label{table:ecl}
\end{planotable}
\end{small}

\subsection{Cepheids}

In Table~\ref{table:ceph} we present the parameters of 52 Cepheids in
the M31F field, sorted by the period $P$.  For each Cepheid we present
its name, J2000.0 coordinates, period $P$, flux-weighted mean
magnitudes $\langle V\rangle$ and (when available) $\langle I\rangle$
and $\langle B\rangle$, and the $V$-band amplitude of the variation
$A$.  In Figure~\ref{fig:ceph} we show the phased $B,V,I$ lightcurves
of our Cepheids. Also shown is the best fit template lightcurve 
(Stetson 1996), which was fitted to the $V$ data and then for the $I$
data only the zero-point offset was allowed. For the $B$-band data,
lacking the template lightcurve parameterization (Stetson 1996), we
used the $V$-band template, allowing for different zero-points and
amplitudes. With our limited amounts of $B$-band data this approach
produces mostly satisfactory results, but extending the
template-fitting approach of Stetson (1996) to the $B$-band (and
possibly other popular bands) would be most useful. 

Some Cepheids seem to be brighter in the $B$ band than in $V$. This
effect is most likely caused by blending, since these variables are
located in regions densely populated by stars, but could partially be
due to blue binary companions of Cepheids (Evans 1994; Evans \&
Udalski 1994).

\begin{figure}[p]
\plotfiddle{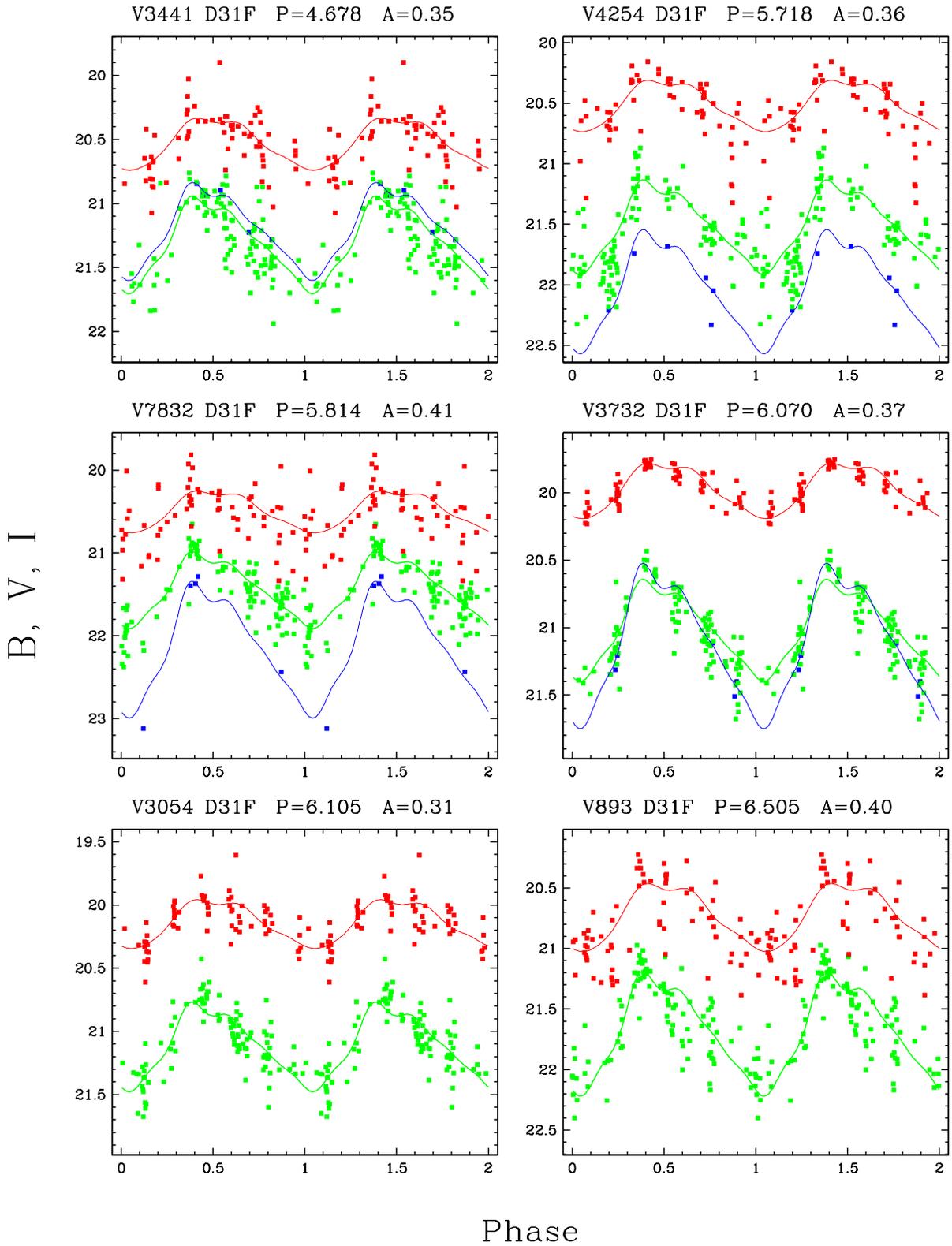}{19.5cm}{0}{83}{83}{-260}{-40}
\caption{$BVI$ lightcurves of Cepheid variables found in the field
M31F. The thin continuous line represents the best fit Cepheid
template for each star and photometric band. $B$ (if present) is
usually the faintest and $I$ (if present) is usually the brightest.}
\label{fig:ceph}
\end{figure}

\addtocounter{figure}{-1}
\begin{figure}[p]
\plotfiddle{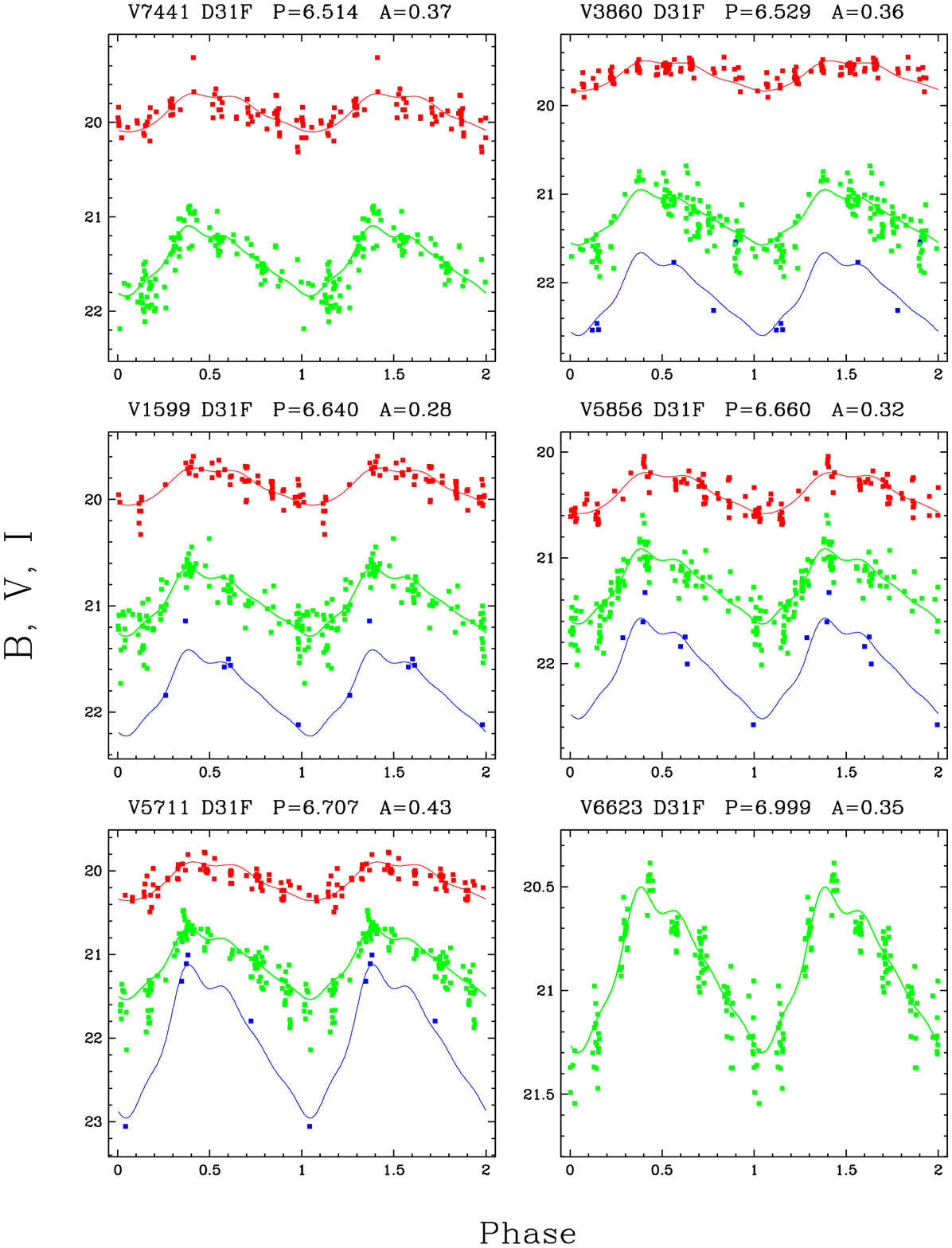}{19.5cm}{0}{83}{83}{-260}{-40}
\caption{Continued.}
\end{figure}

\addtocounter{figure}{-1}
\begin{figure}[p]
\plotfiddle{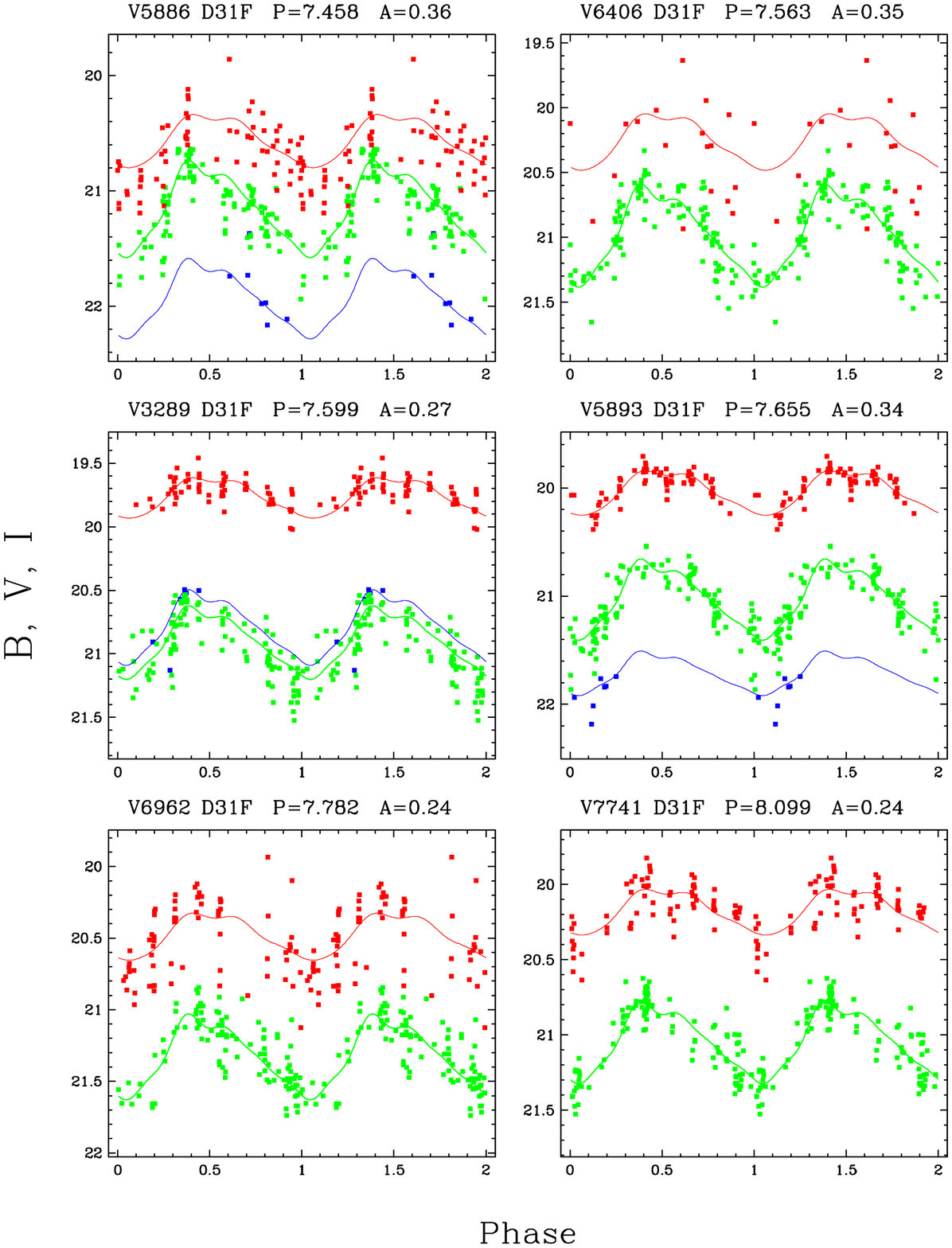}{19.5cm}{0}{83}{83}{-260}{-40}
\caption{Continued.}
\end{figure}

\addtocounter{figure}{-1}
\begin{figure}[p]
\plotfiddle{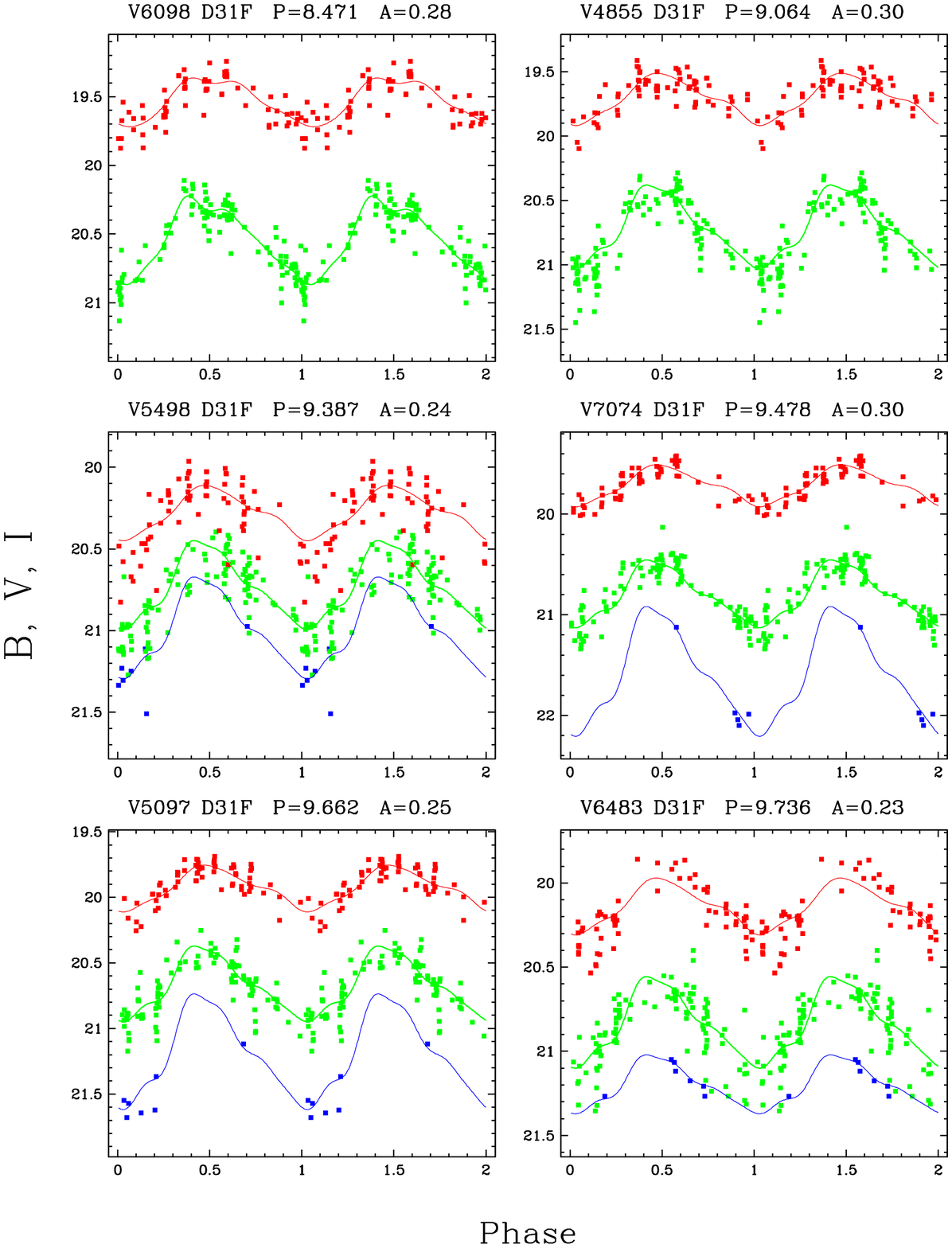}{19.5cm}{0}{83}{83}{-260}{-40}
\caption{Continued.}
\end{figure}

\addtocounter{figure}{-1}
\begin{figure}[p]
\plotfiddle{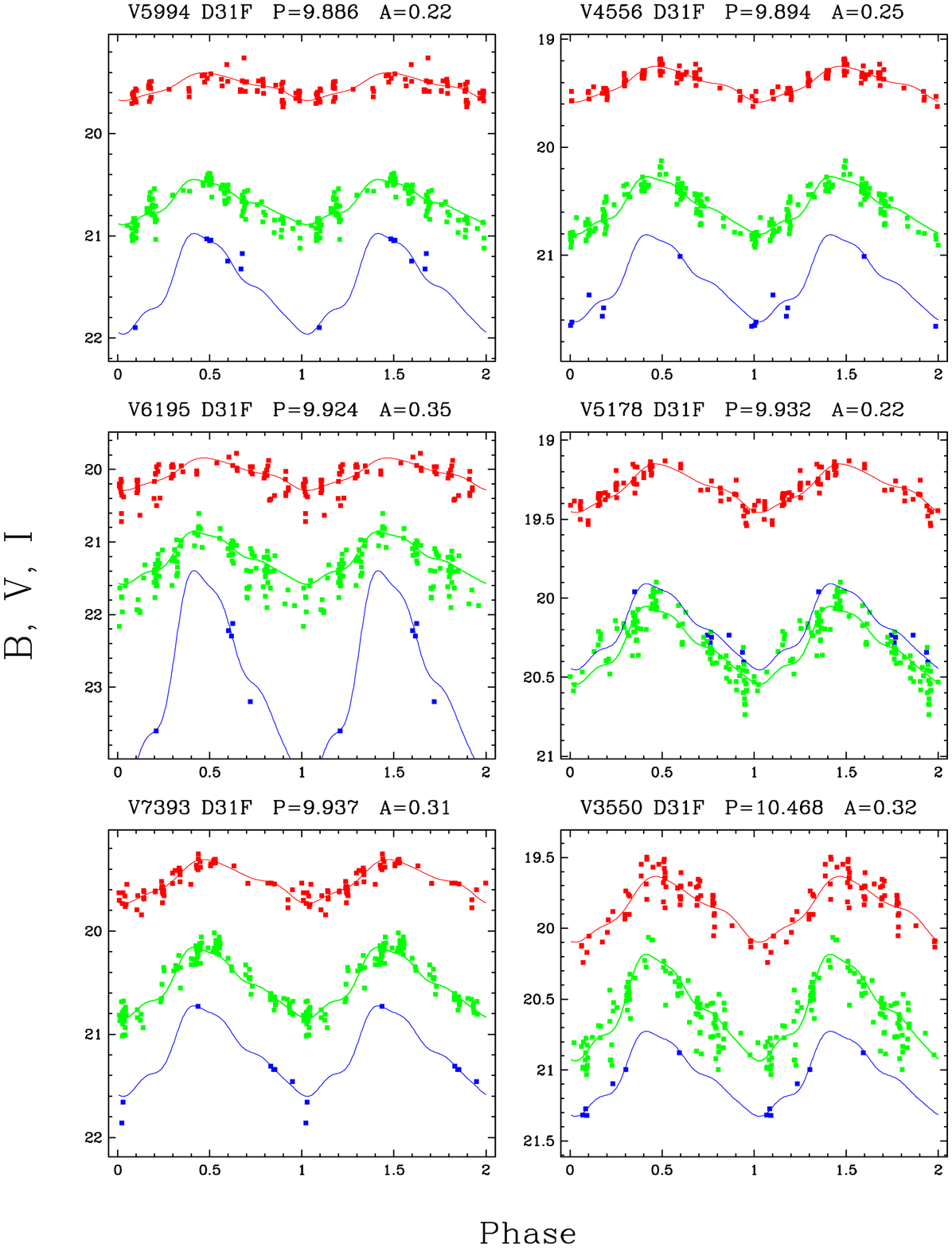}{19.5cm}{0}{83}{83}{-260}{-40}
\caption{Continued.}
\end{figure}

\addtocounter{figure}{-1}
\begin{figure}[p]
\plotfiddle{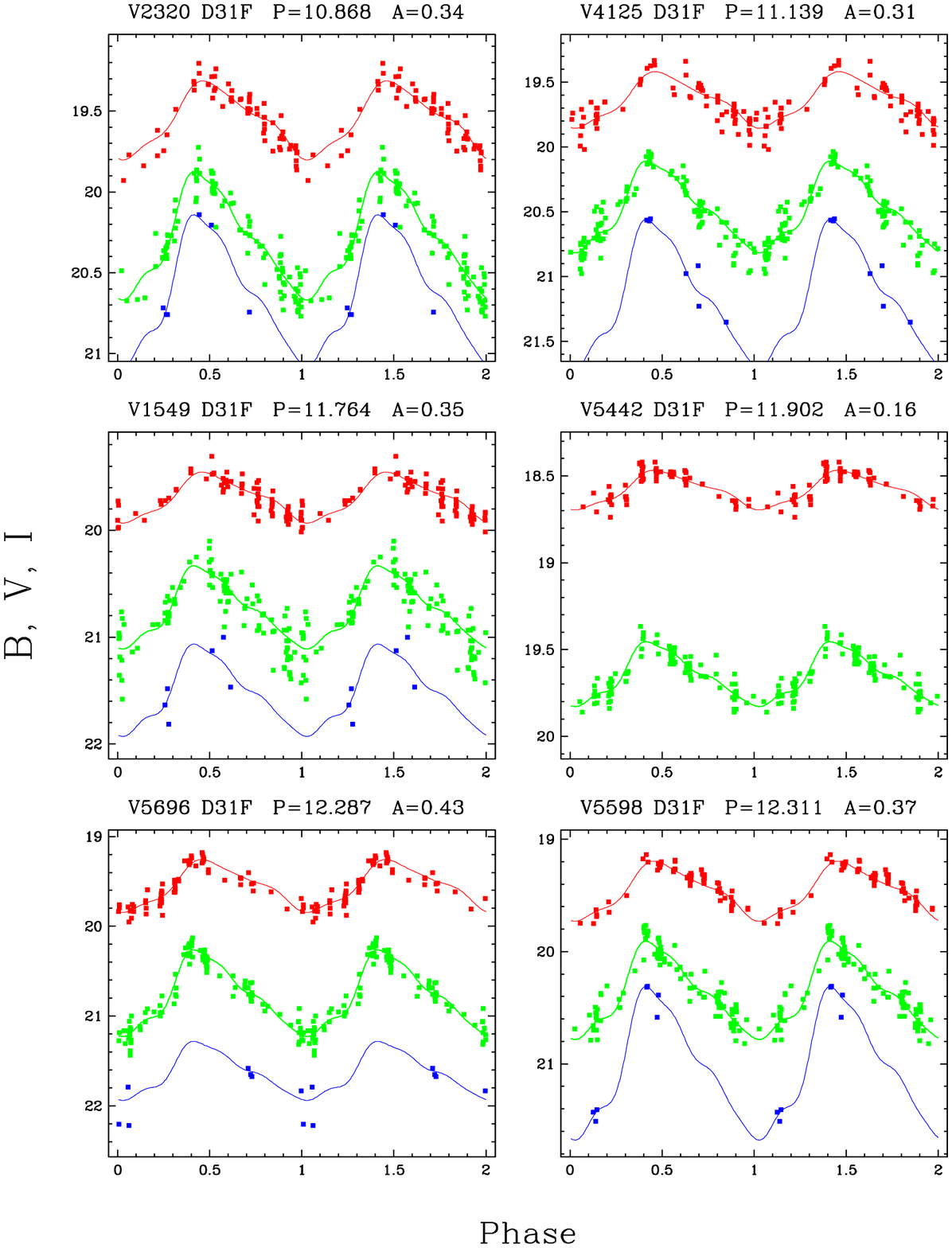}{19.5cm}{0}{83}{83}{-260}{-40}
\caption{Continued.}
\end{figure}

\addtocounter{figure}{-1}
\begin{figure}[p]
\plotfiddle{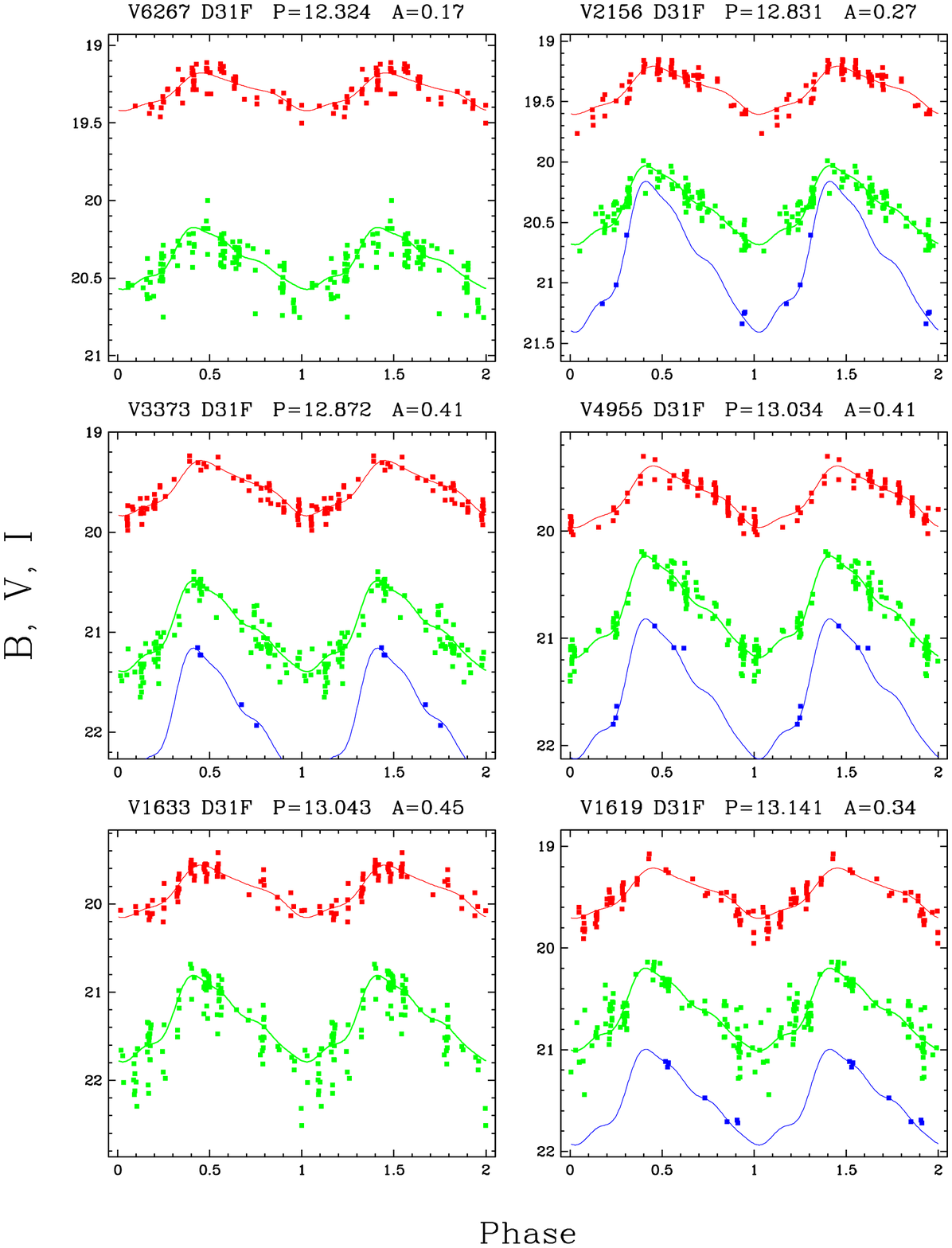}{19.5cm}{0}{83}{83}{-260}{-40}
\caption{Continued.}
\end{figure}

\addtocounter{figure}{-1}
\begin{figure}[p]
\plotfiddle{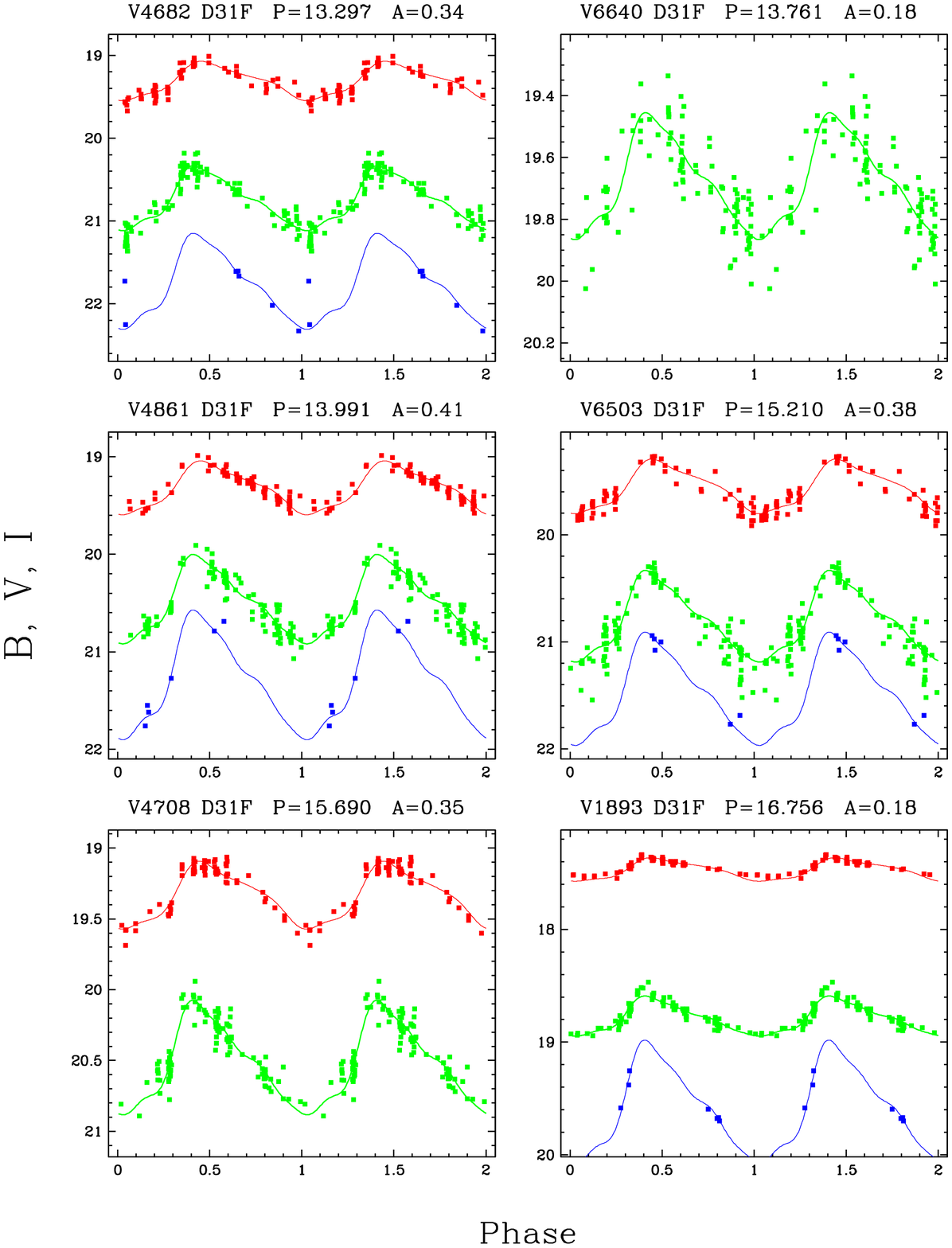}{19.5cm}{0}{83}{83}{-260}{-40}
\caption{Continued.}
\end{figure}

\addtocounter{figure}{-1}
\begin{figure}[p]
\plotfiddle{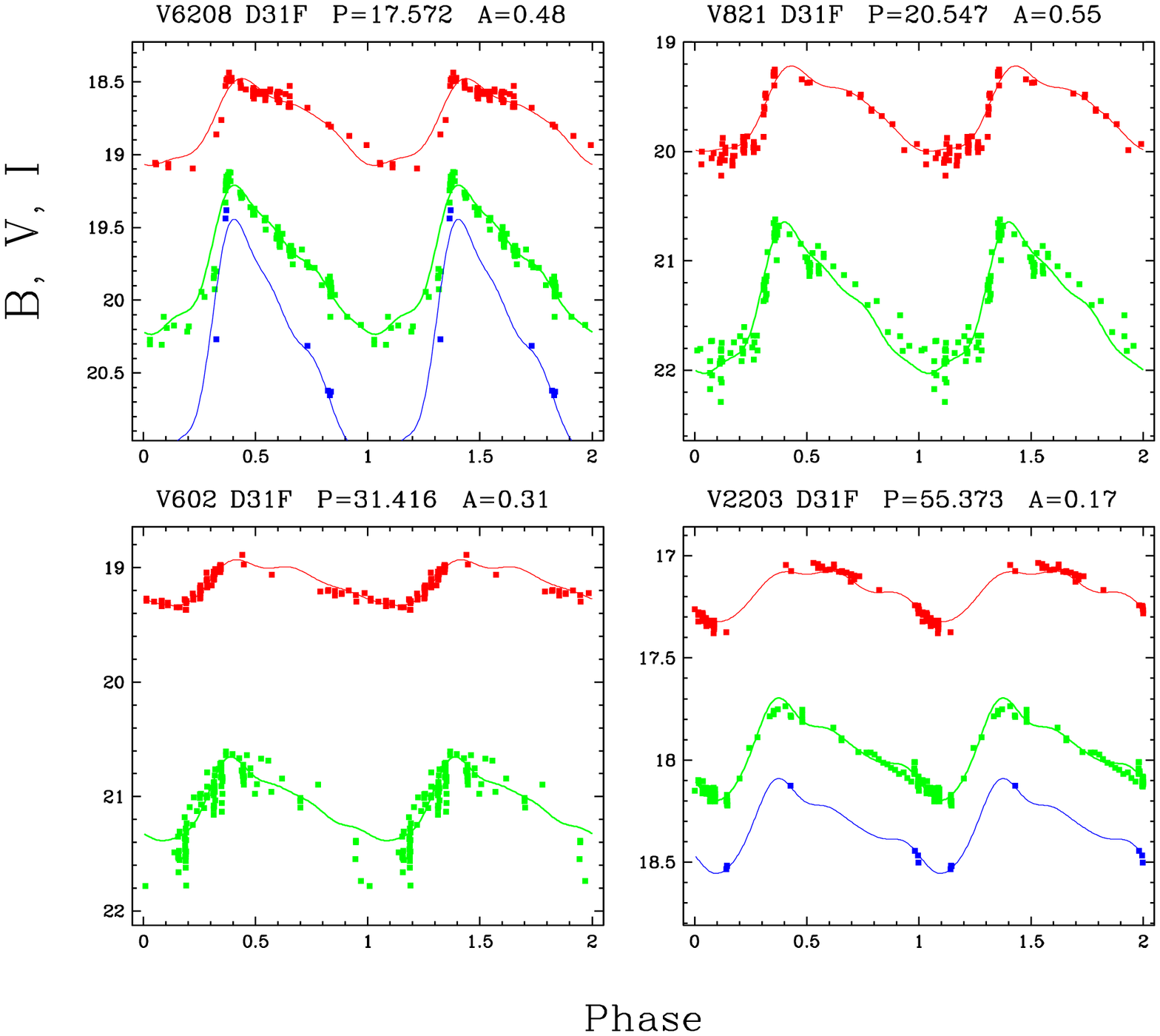}{6.25cm}{0}{83}{83}{-260}{-230}
\caption{Continued.}
\end{figure}
\clearpage

\subsection{Other periodic variables}

\begin{figure}[tp]
\plotfiddle{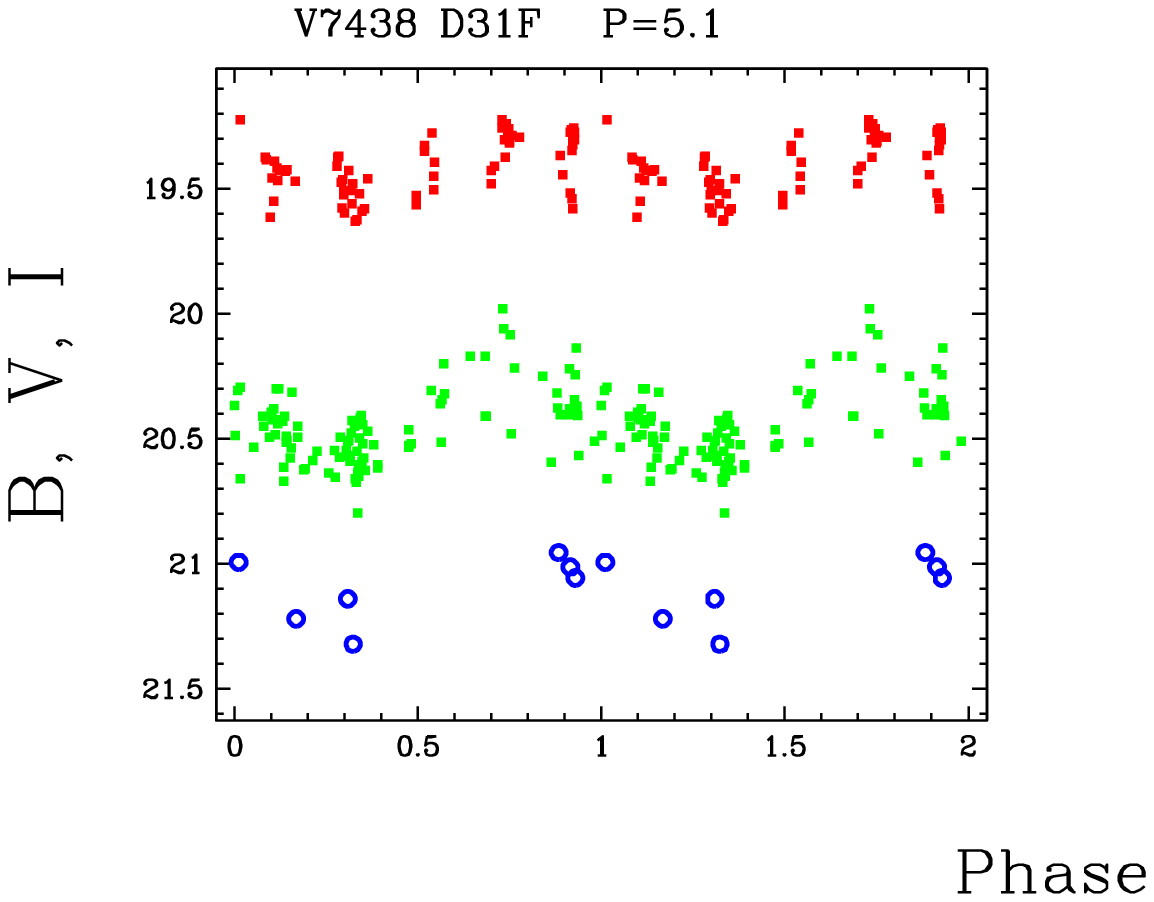}{6.45cm}{0}{83}{83}{-260}{-418}
\caption{$BVI$ lightcurves of the other periodic variable found in the
field M31F.  $B$-band data (shown with the open circles) is the
faintest and $I$ is the brightest.}
\label{fig:per}
\end{figure}

For one of the variables preliminarily classified as an eclipsing
binary we decided upon closer examination to classify it as an ``other
periodic variable''.  In Table~\ref{table:per} we present the
parameters of this possible periodic variable. In Figure~\ref{fig:per}
we show its phased $BVI$ lightcurves.  We present its name, J2000.0
coordinates, period $P$, error-weighted mean magnitudes $\bar{V}$,
$\bar{I}$ and $\bar{B}$. To quantify the amplitude of the variability,
we also give the standard deviations of the measurements in the $BVI$
bands, $\sigma_{V},\sigma_{I}$ and $\sigma_{B}$.

\begin{figure}[p]
\plotfiddle{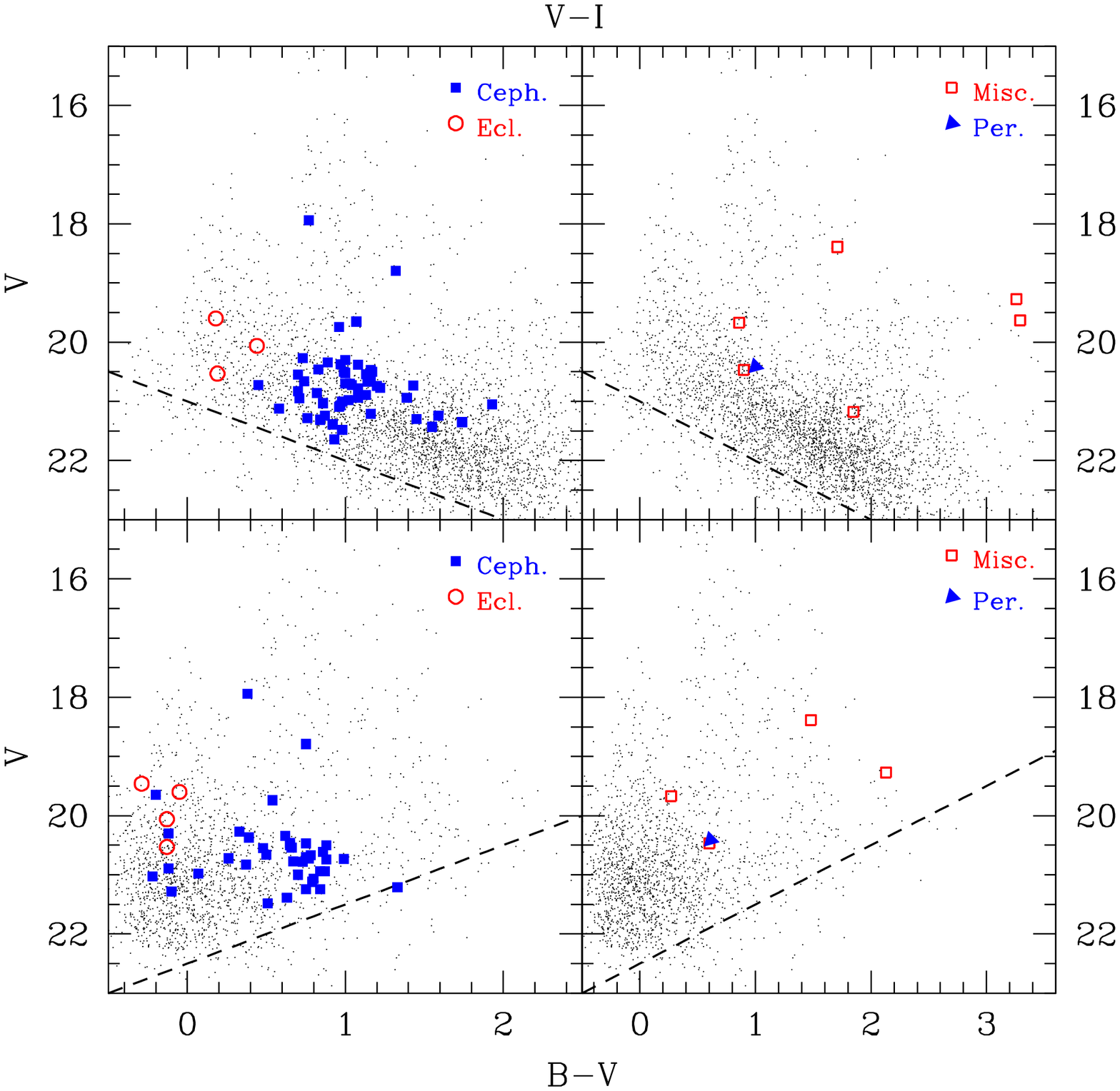}{13cm}{0}{85}{85}{-260}{-135}
\caption{$V,\;V-I$ (upper panels) and $B,\;B-V$ (lower panels)
color-magnitude diagrams for the variable stars found in the field
M31F. The eclipsing binaries and Cepheids are plotted in the left
panels and the other periodic variables and miscellaneous variables
are plotted in the right panels. The dashed lines correspond to the
$I$ detection limit of $I\sim21\;{\rm mag}$ (upper panels) and the $B$
detection limit of $B\sim22.5.\;{\rm mag}$ (lower panels).
\label{fig:cmd}}
\end{figure}

\begin{figure}[t]
\plotfiddle{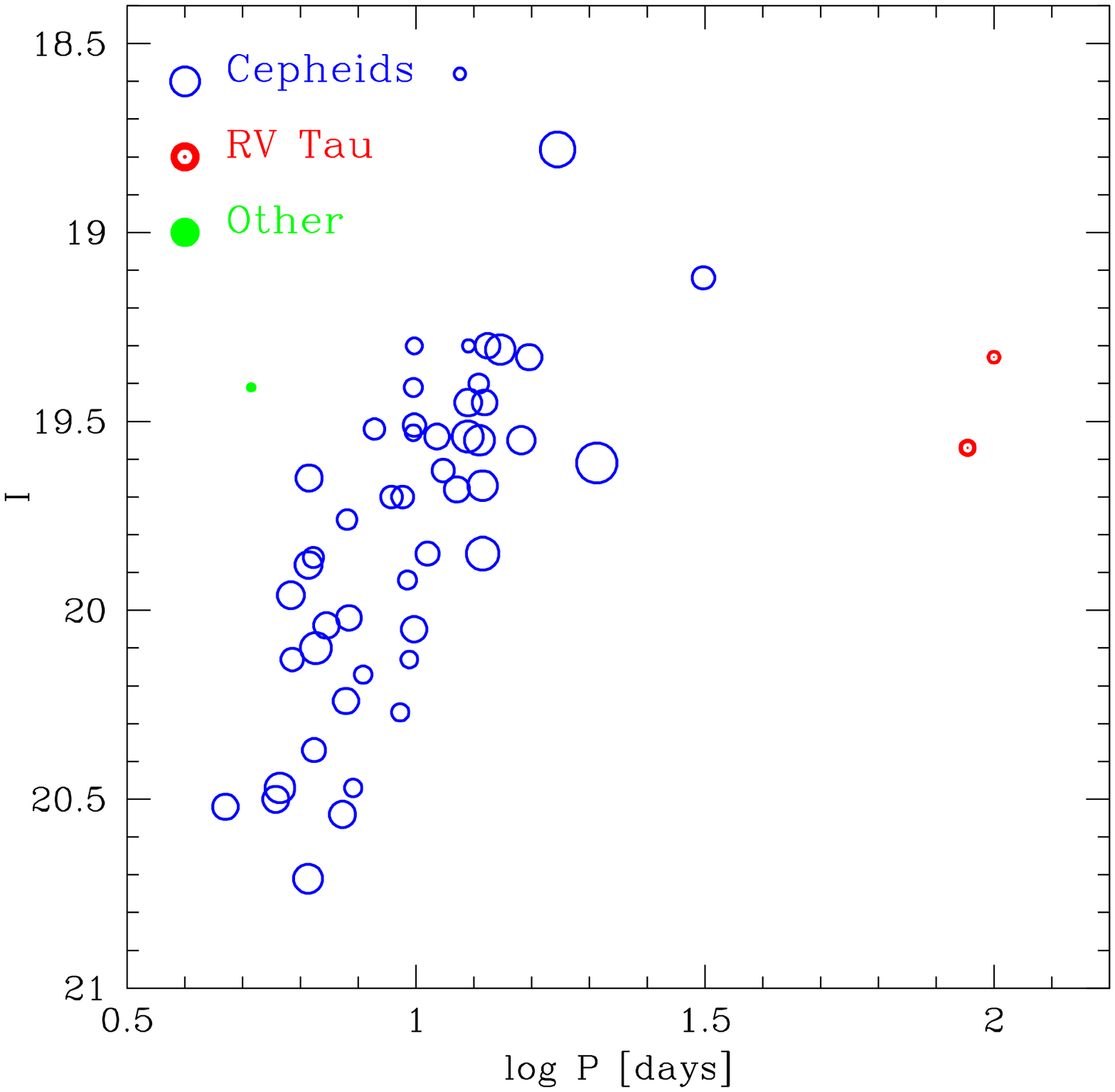}{8cm}{0}{50}{50}{-160}{-85}
\caption{Diagram of $\log{P}$ vs. $I$ for the Cepheids (open circles),
RV Tau (dotted circles) variables and the other periodic variable
(filled circle). The sizes of the circles are proportional to the $V$
amplitude of the variability.}
\label{fig:pl}
\end{figure}

The period of V7438 D31F was taken to be half of the period determined
by fitting a simple eclipsing binary lightcurve, so it should only be
treated as a first approximation of its true value. Inspection of the
$V,V-I$ and $V,B-V$ color-magnitude diagrams (Figure~\ref{fig:cmd})
reveals that the variable lands in the regions occupied by
Cepheids. V7438 D31F has been previously identified by Baade \& Swope
(1965) as a Cepheid with a period of 5.12 days, very close to our
value. In the P-L diagram (Figure~\ref{fig:pl}), however, it is
located above the region occupied by Cepheid variables, indicating
that it is possibly a blend. Another fact which may favor this
explanation is the small amplitude of its variability.

\begin{small}
\tablenum{2}
\begin{planotable}{lrrrrrrr}
\tablewidth{35pc}
\tablecaption{DIRECT Cepheids in M31F}
\tablehead{ \colhead{Name} & \colhead{$\alpha_{J2000.0}$} &
\colhead{$\delta_{J2000.0}$} & \colhead{$P$}  &
\colhead{} & \colhead{} & \colhead{} & \colhead{}\\
\colhead{(D31F)} & \colhead{(deg)} & \colhead{(deg)}
& \colhead{$(days)$} & \colhead{$\langle V\rangle$} & \colhead{$\langle I\rangle$}
& \colhead{$\langle B\rangle$} & \colhead{$A$} }
\startdata   
V3441\ldots   & 10.1201 & 40.7667 &  4.678 & 21.28 & 20.52 & 21.18 & 0.35 \nl 
V4254\dotfill & 10.1111 & 40.6766 &  5.718 & 21.48 & 20.50 & 21.99 & 0.36 \nl 
V7832\dotfill &  9.9963 & 40.7972 &  5.814 & 21.39 & 20.47 & 22.02 & 0.41 \nl 
V3732\dotfill & 10.1189 & 40.6764 &  6.070 & 20.98 & 19.96 & 21.05 & 0.37 \nl 
V3054\dotfill & 10.1243 & 40.7911 &  6.105 & 21.09 & 20.13 &\nodata& 0.31 \nl 
 V893\dotfill & 10.1716 & 40.7771 &  6.505 & 21.64 & 20.71 &\nodata& 0.40 \nl
V7441\dotfill & 10.0247 & 40.6533 &  6.514 & 21.43 & 19.88 &\nodata& 0.37 \nl 
V3860\dotfill & 10.1154 & 40.7284 &  6.529 & 21.24 & 19.65 & 22.08 & 0.36 \nl 
V1599\dotfill & 10.1470 & 40.7687 &  6.640 & 20.94 & 19.86 & 21.78 & 0.28 \nl 
V5856\dotfill & 10.0799 & 40.6725 &  6.660 & 21.24 & 20.37 & 21.99 & 0.32 \nl 
V5711\dotfill & 10.0828 & 40.6802 &  6.707 & 21.07 & 20.10 & 21.87 & 0.43 \nl 
V6623\dotfill & 10.0591 & 40.6814 &  6.999 & 20.86 & 20.04 &\nodata& 0.35 \nl 
V5886\dotfill & 10.0788 & 40.6903 &  7.458 & 21.12 & 20.54 & 21.91 & 0.36 \nl 
V6406\dotfill & 10.0662 & 40.6569 &  7.563 & 20.95 & 20.24 &\nodata& 0.35 \nl 
V3289\dotfill & 10.1232 & 40.7453 &  7.599 & 20.89 & 19.76 & 20.77 & 0.27 \nl 
V5893\dotfill & 10.0797 & 40.6513 &  7.655 & 21.00 & 20.02 & 21.70 & 0.34 \nl 
V6962\dotfill & 10.0462 & 40.6923 &  7.782 & 21.31 & 20.47 &\nodata& 0.24 \nl 
V7741\dotfill & 10.0076 & 40.6885 &  8.099 & 21.03 & 20.17 & 20.81 & 0.24 \nl 
V6098\dotfill & 10.0748 & 40.6395 &  8.471 & 20.52 & 19.52 &\nodata& 0.28 \nl 
V4855\dotfill & 10.0966 & 40.8035 &  9.064 & 20.70 & 19.70 &\nodata& 0.30 \nl 
V5498\dotfill & 10.0883 & 40.6611 &  9.387 & 20.72 & 20.27 & 20.98 & 0.24 \nl 
V7074\dotfill & 10.0388 & 40.7778 &  9.478 & 20.78 & 19.70 & 21.51 & 0.30 \nl 
V5097\dotfill & 10.0967 & 40.6804 &  9.662 & 20.66 & 19.92 & 21.16 & 0.25 \nl 
V6483\dotfill & 10.0637 & 40.6739 &  9.736 & 20.83 & 20.13 & 21.20 & 0.23 \nl 
V5994\dotfill & 10.0774 & 40.6532 &  9.886 & 20.67 & 19.53 & 21.45 & 0.22 \nl 
V4556\dotfill & 10.1054 & 40.7022 &  9.894 & 20.54 & 19.41 & 21.20 & 0.25 \nl 
V6195\dotfill & 10.0722 & 40.6463 &  9.924 & 21.21 & 20.05 & 22.54 & 0.35 \nl 
V5178\dotfill & 10.0949 & 40.6807 &  9.932 & 20.30 & 19.30 & 20.18 & 0.22 \nl 
V7393\dotfill & 10.0278 & 40.6364 &  9.937 & 20.50 & 19.51 & 21.15 & 0.31 \nl 
V3550\dotfill & 10.1188 & 40.7607 & 10.468 & 20.55 & 19.85 & 21.03 & 0.32 \nl 
V2320\dotfill & 10.1341 & 40.7656 & 10.868 & 20.27 & 19.54 & 20.60 & 0.34 \nl 
V4125\dotfill & 10.1109 & 40.7332 & 11.139 & 20.46 & 19.63 & 21.11 & 0.31 \nl 
V1549\dotfill & 10.1488 & 40.7569 & 11.764 & 20.72 & 19.68 & 21.49 & 0.35 \nl 
V5442\dotfill & 10.0892 & 40.6766 & 11.902 & 19.65 & 18.58 & 19.45 & 0.16 \nl 
V5696\dotfill & 10.0806 & 40.7520 & 12.287 & 20.74 & 19.54 & 21.62 & 0.43 \nl 
V5598\dotfill & 10.0855 & 40.6741 & 12.311 & 20.34 & 19.45 & 20.96 & 0.37 \nl 
V6267\dotfill & 10.0684 & 40.6995 & 12.324 & 20.38 & 19.30 & 19.27 & 0.17 \nl 
V2156\dotfill & 10.1353 & 40.7925 & 12.831 & 20.37 & 19.40 & 20.76 & 0.27 \nl 
V3373\dotfill & 10.1199 & 40.7971 & 12.872 & 20.94 & 19.55 & 21.81 & 0.41 \nl 
V4955\dotfill & 10.0959 & 40.7808 & 13.034 & 20.70 & 19.67 & 21.45 & 0.41 \nl 
V1633\dotfill & 10.1461 & 40.7760 & 13.043 & 21.30 & 19.85 &\nodata& 0.45 \nl 
V1619\ldots   & 10.1503 & 40.6545 & 13.141 & 20.61 & 19.45 & 21.47 & 0.34 \nl 
V4682\dotfill & 10.1046 & 40.6552 & 13.297 & 20.73 & 19.30 & 21.72 & 0.34 \nl 
V6640\dotfill & 10.0593 & 40.6558 & 13.761 & 19.67 &\nodata&\nodata& 0.18 \nl 
V4861\dotfill & 10.0972 & 40.7849 & 13.991 & 20.47 & 19.31 & 21.22 & 0.41 \nl 
V6503\dotfill & 10.0615 & 40.7215 & 15.210 & 20.77 & 19.55 & 21.44 & 0.38 \nl 
V4708\dotfill & 10.1039 & 40.6631 & 15.690 & 20.50 & 19.33 & 21.38 & 0.35 \nl 
V1893\dotfill & 10.1415 & 40.7417 & 16.756 & 18.79 & 17.47 & 19.54 & 0.18 \nl 
V6208\dotfill & 10.0714 & 40.6561 & 17.572 & 19.74 & 18.78 & 20.28 & 0.48 \nl 
 V821\dotfill & 10.1760 & 40.7640 & 20.547 & 21.35 & 19.61 &\nodata& 0.55 \nl 
 V602\dotfill & 10.1880 & 40.7392 & 31.416 & 21.05 & 19.12 &\nodata& 0.31 \nl 
V2203\dotfill & 10.1369 & 40.7306 & 55.373 & 17.94 & 17.17 & 18.32 & 0.17 \nl 
\enddata
\label{table:ceph}
\end{planotable}
\end{small}

\subsection{Miscellaneous  variables}	
	
In Table~\ref{table:misc} we present the parameters of seven
miscellaneous variables in the M31F field, sorted by increasing value
of the mean magnitude $\bar{V}$.  In Figure~\ref{fig:misc} we show the
unphased $VI$ lightcurves of the miscellaneous variables. For each
variable we present its name, J2000.0 coordinates and mean magnitudes
$\bar{V}, \bar{I}$ and $\bar{B}$.  To quantify the amplitude of the
variability, we also give the standard deviations of the measurements
in $BVI$ bands, $\sigma_{V}, \sigma_{I}$ and $\sigma_{B}$.  In the
``Comments'' column we give a rather broad sub-classification of the
variability.

All of the variables seem to represent the LP type of variability. A
closer inspection of the color-magnitude diagrams
(Figure~\ref{fig:cmd}) reveals that three variables (V667, V1229 and
V2285 D31F) land in the same area as Cepheids. Based on their
lightcurves it was possible to roughly estimate the periods of the
first two to be around 90 and 100 days, respectively. Using these
periods to place the stars on the P-L diagram (Figure~\ref{fig:pl})
suggests they may be RV Tauri type variables. V1665 and V1724 D31F are
most likely Mira-type variables, based on their location in the
color-magnitude diagrams.

\begin{small}
\tablenum{3}
\begin{planotable}{cccrccccccl}
\tablewidth{40pc}
\tablecaption{DIRECT Other Periodic Variables in M31F}
\tablehead{ \colhead{Name} & \colhead{$\alpha_{J2000.0}$} &
\colhead{$\delta_{J2000.0}$} & \colhead{$P$} &
\colhead{} & \colhead{} & \colhead{} & \colhead{} & \colhead{} & \colhead{} & \colhead{} \\
\colhead{(D31F)} &  \colhead{(deg)} &  \colhead{(deg)} &
\colhead{$(days)$} & \colhead{$\bar{V}$} &
\colhead{$\bar{I}$} & \colhead{$\bar{B}$} & \colhead{$\sigma_V$} &
\colhead{$\sigma_I$} & \colhead{$\sigma_B$} & \colhead{Comments} }
\startdata   
V7438\ldots   & 10.0252 & 40.6442 &   5.1 & 20.41 & 19.41 & 21.02 & 0.15 & 0.11 & 0.13 & Cepheid?\nl 
\enddata
\label{table:per}
\tablecomments{V7438 was identified by Baade \& Swope (1965) as Cepheid
variable 232 with $P=5.12$ days.}
\end{planotable}
\end{small}

\begin{small}
\tablenum{4}
\begin{planotable}{llllllllll}
\tablewidth{35pc}
\tablecaption{DIRECT Miscellaneous Variables in M31F}
\tablehead{ \colhead{Name} & \colhead{$\alpha_{J2000.0}$} &
\colhead{$\delta_{J2000.0}$} & \colhead{} & \colhead{} &
\colhead{} & \colhead{} & \colhead{} & \colhead{} & \colhead{} \\
\colhead{(D31F)} & \colhead{(deg)} & \colhead{(deg)} &
\colhead{$\bar{V}$} & \colhead{$\bar{I}$} & \colhead{$\bar{B}$} &
\colhead{$\sigma_V$} & \colhead{$\sigma_I$} & \colhead{$\sigma_B$} &
\colhead{Comments} }
\startdata   
 V244\ldots   & 10.2103 & 40.7380 & 18.39 & 16.68 & 19.87 & 0.12 & 0.07 & 0.05 & LP\nl 
V1665\dotfill & 10.1460 & 40.7562 & 19.27 & 16.01 & 21.40 & 0.20 & 0.16 & 0.10 & LP\nl 
V1724\dotfill & 10.1446 & 40.7499 & 19.63 & 16.34 &  0.00 & 0.24 & 0.13 & 0.00 & LP\nl 
V2285\dotfill & 10.1373 & 40.6841 & 19.67 & 18.81 & 19.94 & 0.08 & 0.07 & 0.08 & RV Tau?\nl 
V1229\dotfill & 10.1590 & 40.7389 & 20.47 & 19.57 & 21.07 & 0.23 & 0.10 & 0.18 & RV Tau?\nl
 V764\dotfill & 10.1775 & 40.8110 & 20.51 &  0.00 &  0.00 & 0.35 & 0.00 & 0.00 & LP\nl 
 V667\dotfill & 10.1852 & 40.7265 & 21.18 & 19.33 &  0.00 & 0.35 & 0.22 & 0.00 & RV Tau?\nl 
\enddata
\label{table:misc}
\end{planotable}
\end{small}

\subsection{Comparison with other catalogs}

The area of the M31F field coincides with two overlapping fields observed
by Baade. The catalogs of variable stars discovered in those fields are
given by Gaposchkin (1962, field II) and Baade \& Swope (1965, field III).
We succeeded in the cross-identification of all but one of the 55 Cepheid
variables found in field III with stars on our template. We have discovered
independently 24 of those Cepheids and found a very good agreement between
the period determinations. We have also confirmed the periods of 23 other
Cepheids, which eluded our detection, in large part due to their faintness
and the strict criteria we have imposed in our process of Cepheid selection
(see Table~\ref{table:crossid} for cross-ids).

Out of the 38 unique Cepheid variables listed in the field II catalog,
located within our M31F field, we have found 14 Cepheids and confirmed the
periods of two. The remaining field II Cepheids have evaded positive
cross-identification with our template stars. 

Another overlapping variable star catalog is given by Magnier et
al.~(1997, hereafter Ma97). Out of the three variable stars in Ma97
which are in our M31F field, we cross-identified one, also classifying
it as a Cepheid. The other two did not qualify as variable star
candidates because of low $J_S$ values.

\begin{small}
\tablenum{5}
\begin{planotable}{lrrrrrrr}
\tablewidth{35pc}
\tablecaption{Cross-Identifications of the DIRECT Cepheid Variables in M31F}
\tablehead{ \colhead{Name} & \colhead{$P$} &\colhead{} &
\colhead{$P$} & \colhead{} & \colhead{$P$} & \colhead{} &\colhead{$P$}\\
\colhead{(D31F)} & \colhead{$(days)$} & \colhead{field II} &
\colhead{($days$)} & \colhead{field III} &  \colhead{$(days)$} &
\colhead{Other} & \colhead{$(days)$}} 
\startdata   
V3441\ldots   &  4.678 &     &        & 108 &  5.000 & &  \nl
V4254\dotfill &  5.718 & 351 &  5.719 & 153 &  5.721 & &  \nl
V3732\dotfill &  6.070 & 352 &  6.071 & 175 &  6.074 & &  \nl
V3054\dotfill &  6.105 &     &        &  92 &  6.101 & &  \nl
V7441\dotfill &  6.514 & 230 &  6.508 &     &        & &  \nl
V3860\dotfill &  6.529 &     &        &  24 &  6.526 & &  \nl
V1599\dotfill &  6.640 &     &        & 145 &  6.637 & &  \nl
V5856\dotfill &  6.660 & 328 &  6.660 &  30 &  6.700 & &  \nl
V5711\dotfill &  6.707 & 330 &  6.709 &  29 &  6.708 & &  \nl
V6623\dotfill &  6.999 & 326 &  7.000 &     &        & &  \nl
V5886\dotfill &  7.458 & 332 &  7.457 &  70 &  7.463 & &  \nl
V6406\dotfill &  7.563 & 319 &  7.553 &     &        & &  \nl
V5893\dotfill &  7.655 & 320 &  7.823 &     &        & &  \nl
V6962\dotfill &  7.782 & 225 &  7.780 &     &        & &  \nl
V7741\dotfill &  8.099 & 222 &  8.094 &     &        & &  \nl
V6098\dotfill &  8.471 & 315 &  8.464 &     &        & &  \nl
V4855\dotfill &  9.064 &     &        &  51 &  9.085 & &  \nl
V5097\dotfill &  9.662 & 348 &  9.662 &  68 &  9.678 & &  \nl
V6483\dotfill &  9.736 & 325 &  9.483 &     &        & &  \nl
V4556\dotfill &  9.894 & 341 &  9.881 &  60 &  9.870 & &  \nl
V6195\dotfill &  9.924 & 318 &  9.921 &     &        & &  \nl
V7393\dotfill &  9.937 & 234 &  9.933 &     &        & Ma97 4& 9.0 \nl
V3550\dotfill & 10.468 &     &        &  72 & 10.461 & &  \nl
V2320\dotfill & 10.868 &     &        & 109 & 10.858 & &  \nl
V4125\dotfill & 11.139 &     &        &  20 & 11.147 & &  \nl
V1549\dotfill & 11.764 &     &        &  54 & 11.766 & &  \nl
V5696\dotfill & 12.287 &     &        &  17 & 12.286 & &  \nl
V5598\dotfill & 12.311 & 329 & 12.312 & 135 & 12.358 & &  \nl
V6267\dotfill & 12.324 & 334 & 12.294 & 133 & 12.284 & &  \nl
V2156\dotfill & 12.831 &     &        & 128 & 12.821 & &  \nl
V4955\dotfill & 13.034 &     &        &  14 & 13.051 & &  \nl
V1633\dotfill & 13.043 &     &        & 107 & 13.021 & &  \nl
V1619\dotfill & 13.141 & 415 & 13.125 &  31 &  0.000 & &  \nl
V4682\dotfill & 13.297 & 357 & 13.293 &     &        & &  \nl
V4861\dotfill & 13.991 &     &        &  74 & 13.966 & &  \nl
V6503\dotfill & 15.210 & 339 & 15.232 & 150 & 15.216 & &  \nl
V4708\dotfill & 15.690 & 355 & 15.699 & 114 & 15.625 & &  \nl
V6208\dotfill & 17.572 &     & 17.569 &     &        & H22 &  17.60 \nl
\enddata
\label{table:crossid}
\tablecomments{
field II refers to the catalog published by Gaposchkin (1962); field
III - Baade \& Swope (1965); Ma97 - Magnier et al. (1997); H - Hubble(1929)}
\end{planotable}
\end{small}

\begin{figure}[p]
\plotfiddle{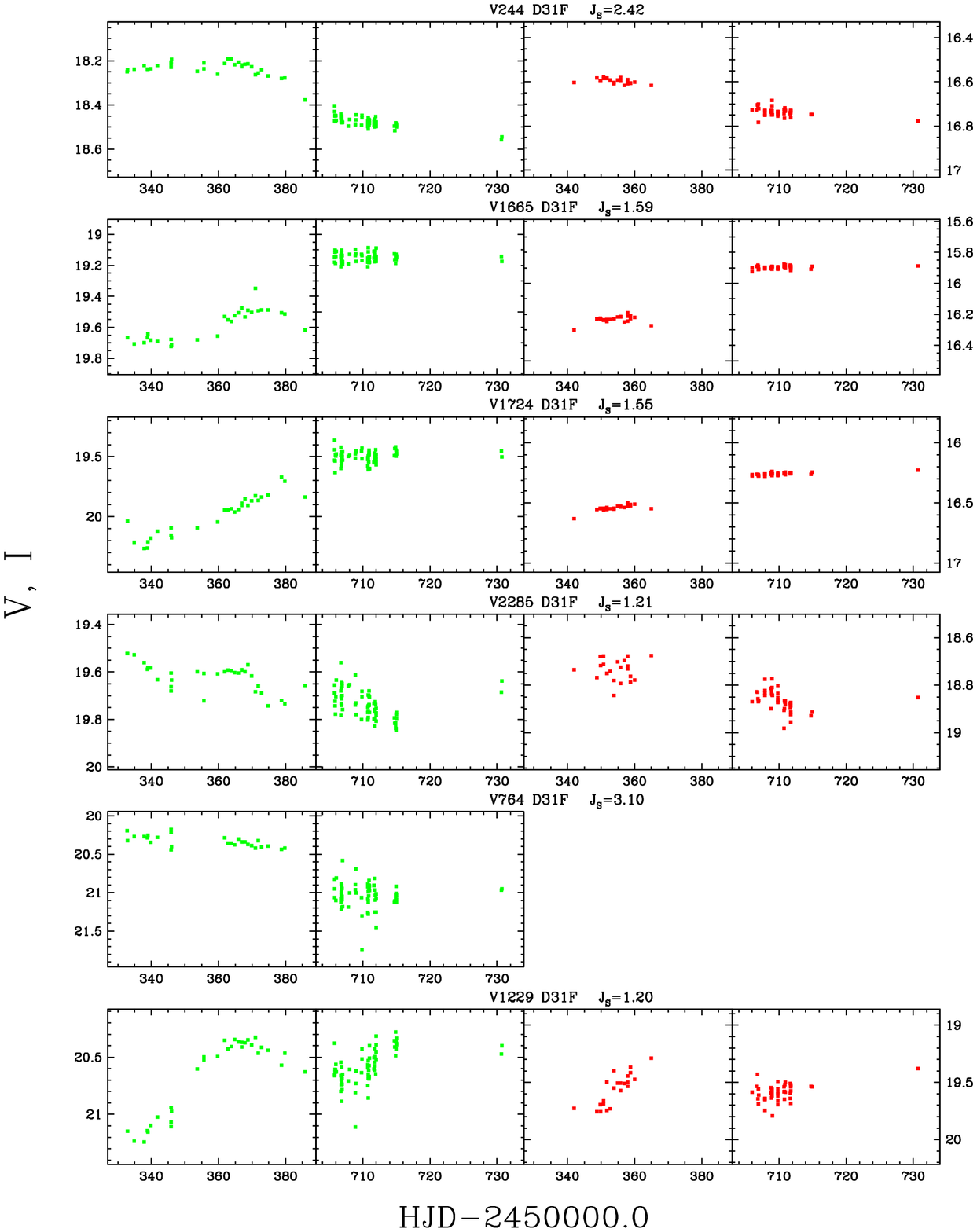}{19.5cm}{0}{83}{83}{-260}{-40}
\caption{$VI$ lightcurves of the miscellaneous variables found in the
field M31F.  $I$ (if present) is plotted in the two right panels.
$B$-band data is not shown.}
\label{fig:misc}
\end{figure}

\addtocounter{figure}{-1}
\begin{figure}[p]
\plotfiddle{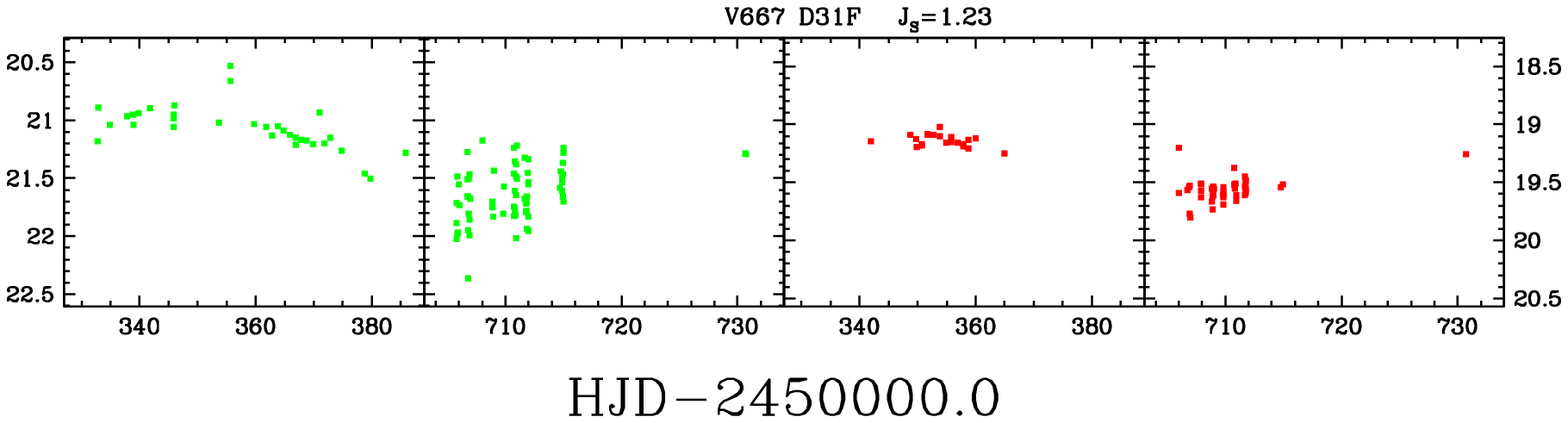}{2cm}{0}{83}{83}{-260}{-495}
\caption{Continued.}
\end{figure}

\section{Discussion}

\begin{figure}[t]   
\plotfiddle{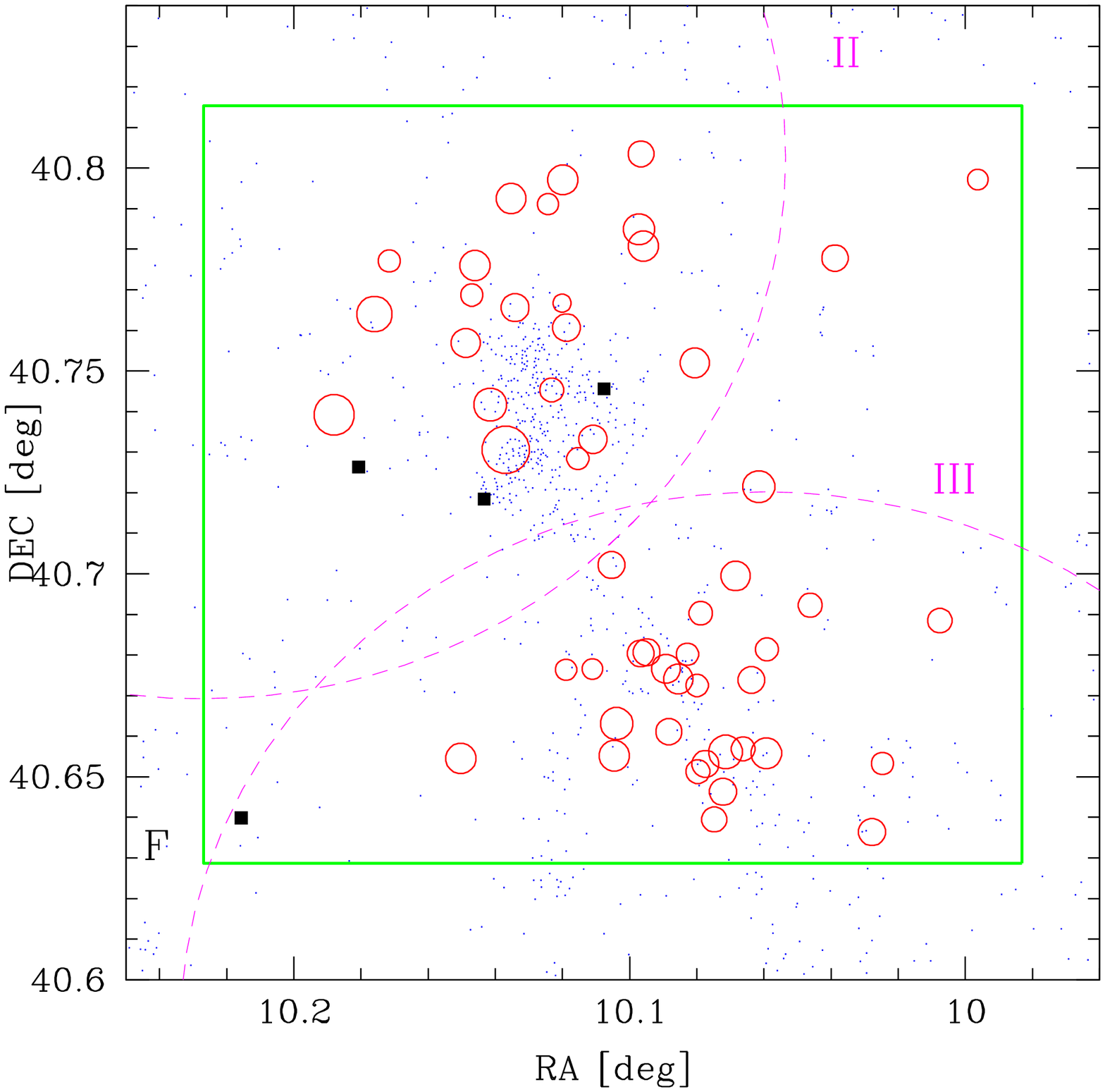}{8cm}{0}{50}{50}{-160}{-85}
\caption{Location of eclipsing binaries (filled squares) and Cepheids
(open circles) in the field M31F, along with the blue stars
($B-V<0.4$) selected from the photometric survey of M31 by Magnier et
al.~(1992) and Haiman et al.~(1994). The sizes of the circles
representing the Cepheids variables are proportional to the logarithm
of their period. Fields II and III observed by Baade are marked with
dashed lines.
\label{fig:xy}}
\end{figure}

In Figure~\ref{fig:cmd} we show $V,\;V-I$ and $V,\;B-V$
color-magnitude diagrams for the variable stars found in the field
M31F. The eclipsing binaries and Cepheids are plotted in the left
panels and the other periodic variables and miscellaneous variables
are plotted in the right panels. As expected, the eclipsing binaries
occupy the blue upper main sequence of M31 stars. The Cepheid
variables group near $B-V\sim1.0$, with considerable scatter probably
due to the differential reddening across the field. The other periodic
variable is located on the CMD in the part occupied by Cepheids. The
miscellaneous variables are scattered throughout the CMDs and
represent several classes of variability. Two of them are very red
with $V-I>2.0$, and are probably Mira variables.

In Figure~\ref{fig:xy} we plot the location of eclipsing binaries and
Cepheids in the field M31F, along with the blue stars ($B-V<0.4$)
selected from the photometric survey of M31 by Magnier et al.~(1992)
and Haiman et al.~(1994). The sizes of the circles representing the
Cepheids variables are proportional to the logarithm of their
period. As could have been expected, both types of variables group
along the spiral arms, as they represent relatively young populations
of stars. Many Cepheid variables are located in the star-forming
region NGC206.  We will explore various properties of our sample of
Cepheids in the future paper (Sasselov et al.~1999, in preparation).

\acknowledgments{We would like to thank the TAC of the F.~L.~Whipple
Observatory (FLWO) and the TAC of the Michigan-Dartmouth-MIT (MDM)
Observatory for the generous amounts of telescope time devoted to this
project. We are grateful to Bohdan Paczy\'nski for motivating us to
undertake this project and always helpful comments and suggestions.
We thank Lucas Macri for taking some of the data described in this
paper, Przemek Wo\'zniak for his FITS-manipulation programs and Eugene
Magnier for the Cepheid catalogs.  The staff of the MDM and the FLWO
observatories is thanked for their support during the long observing
runs.  KZS was supported by the Harvard-Smithsonian Center for
Astrophysics Fellowship. JK was supported by NSF grant AST-9528096 to
Bohdan Paczy\'nski and by the Polish KBN grant 2P03D011.12.}

\end{document}